\documentclass[sigconf, authorversion]{acmart}

\usepackage{amsmath,amsfonts}
\usepackage{algorithmic}
\usepackage{graphicx}
\usepackage{textcomp}
\usepackage{xcolor}
\usepackage{acronym}
\usepackage{caption}
\usepackage{subcaption}
\usepackage{bbding} 
\usepackage{acronym}
\usepackage{amsthm}
\usepackage{balance}
\usepackage{url}

\newtheorem{definition}{Definition}

\acrodef{GPS}{Global Positioning System}
\acrodef{CNN}{Convolutional Neural Network}
\acrodef{DAT}{Day-After-Tomorrow}
\acrodef{RF}{Radio Frequency}
\acrodef{AI}{Artificial Intelligence}
\acrodef{ML}{Machine Learning}
\acrodef{DL}{Deep Learning}
\acrodef{SDR}{Software-Defined Radio}
\acrodef{BPSK}{Binary Phase Shift Keying}
\acrodef{QPSK}{Quadrature Phase Shift Keying}
\acrodef{mle}{maximum likelihood estimation}
\acrodef{RNN}{Residual Neural Networks}
\acrodef{BLE}{Bluetooth Low Energy}
\acrodef{LoRa}{Long Range}
\acrodef{IoT}{Internet of Things}
\acrodef{RFF}{Radio Frequency Fingerprinting}

\copyrightyear{2023}
\acmYear{2023}
\setcopyright{rightsretained}
\acmConference[ACSAC '23]{Annual Computer Security Applications
Conference}{December 4--8, 2023}{Austin, TX, USA}
\acmBooktitle{Annual Computer Security Applications Conference (ACSAC '23),
December 4--8, 2023, Austin, TX, USA}
\acmDOI{10.1145/3627106.3627192}
\acmISBN{979-8-4007-0886-2/23/12}

\AtBeginDocument{%
  }

\begin{document}

\title{The Day-After-Tomorrow:\\ On the Performance of Radio Fingerprinting over Time}

\author{Saeif AlHazbi}
\email{salhazbi@hbku.edu.qa}
\affiliation{%
  \institution{Division of Information and Computing Technology, College of Science and Engineering, Hamad Bin Khalifa University}
  \city{Doha}  
  \country{Qatar}  
}
\author{Savio Sciancalepore}
\email{s.sciancalepore@tue.nl}
\affiliation{%
  \institution{Eindhoven University of Technology}
  \city{Eindhoven}  
  \country{Netherlands}  
}
\author{Gabriele Oligeri}
\email{goligeri@hbku.edu.qa}
\affiliation{%
  \institution{Division of Information and Computing Technology, College of Science and Engineering, Hamad Bin Khalifa University}
  \city{Doha}  
  \country{Qatar}  
}

\begin{abstract}
The performance of Radio Frequency (RF) Fingerprinting (RFF) techniques is negatively impacted when the training data is not temporally close to the testing data. This can limit the practical implementation of physical-layer authentication solutions. To circumvent this problem, current solutions involve collecting training and testing datasets at close time intervals---this being detrimental to the real-life deployment of any physical-layer authentication solution. We refer to this issue as the Day-After-Tomorrow (DAT) effect, being widely attributed to the temporal variability of the wireless channel, which masks the physical-layer features of the transmitter, thus impairing the fingerprinting process.

In this work, we investigate the DAT effect shedding light on its root causes. Our results refute previous knowledge by demonstrating that the DAT effect is not solely caused by the variability of the wireless channel. Instead, we prove that it is also due to the power cycling of the radios, i.e., the turning off and on of the radios between the collection of training and testing data. We show that state-of-the-art RFF solutions double their performance when the devices under test are not power cycled, i.e., the accuracy increases from about 0.5 to about 1 in a controlled scenario. 
Finally, we show how to mitigate the DAT effect in real-world scenarios, through pre-processing of the I-Q samples. Our experimental results show a significant improvement in accuracy, from approximately 0.45 to 0.85. Additionally, we reduce the variance of the results, making the overall performance more reliable. 
\end{abstract} 




\settopmatter{printfolios=true}

\maketitle

\section{Introduction}
\label{sec:introduction}
\ac{RFF} techniques have attracted the attention of researchers working in the wireless security domain, as a way of authenticating wireless transmitters by identifying unique patterns from received signals~\cite{soltanieh2020_jrid}.  Indeed, \ac{RFF} promises an effective and efficient way to authenticate the transmitting source without involving any crypto technique, thus being particularly suitable for scenarios where devices are characterized by strict battery constraints and high exposure to spoofing attacks. The transmitter does not require any additional computation or transmission, while the authentication process is completely offloaded to the receiver side, i.e., the receiver matches a pre-trained model of known transmitters to patterns extracted from the received signals, thus being able to authenticate the source. Such patterns are indeed unique, being the side-effect of unwanted and unpredictable phenomena such as manufacturing inaccuracies during the production process at the sub-millimetre level and electronic components' impurities. Such (small) differences eventually affect radio signals, which are detectable by receivers that combine \ac{SDR} capabilities with advanced \ac{AI} tools, such as the ones based on \ac{ML} and \ac{DL}~\cite{jagannath2022_arxiv}. 
At the time of this writing, extensive research is available on \ac{RFF}. Some of the contributions focused on wireless communication technologies such as LTE~\cite{abbas2021_nca}, WiFi~\cite{jian2021_tmc}, Zigbee~\cite{bihl2016_tifs}, Bluetooth~\cite{ali2019_access}, LoRa~\cite{shen2022towards}, and ADS-B~\cite{jian2020deep}, to name a few. Other works investigated the suitability of several \ac{AI}-based techniques for addressing the RFF problem, either re-adapting well-known neural networks or casting ad-hoc solutions tailored to the specific technologies and data~\cite{wang2022_tccn},~\cite{zhang2022_commlett},~\cite{gong2020_tifs}. At the same time, a few works highlighted reliability issues pointing out phenomena that prevent the real-life deployment of such techniques. These works include the challenges of carrying out reliable training~\cite{hamdaoui2022}, use of multiple customized signal processing techniques by the devices under test~\cite{brik2008}, nonlinear characteristics of the power amplifiers making the fingerprint unpredictably dependent on the transmission power~\cite{kwon2010}, interplay with heat dissipation, operations in different temperature conditions~\cite{polak2015}, aging of the devices and, last but not least, variable channel conditions~\cite{alshawabka2020}.
In this context, some recent authoritative scientific contributions, such as~\cite{alshawabka2020} and~\cite{hamdaoui2022}, found that training RFF models on one day and testing on another day produces very poor performance, with a drop in the achieved classification accuracy of about $0.5$. In this paper, we refer to such a phenomenon as the \ac{DAT} effect. Note that such studies adopt state-of-the-art classifiers based on the \ac{CNN} \emph{Resnet-50}, re-adapted to accept raw physical-layer signals (i.e., I-Q samples) as the input sequence, where the considered input size was either $N \times 1$ or $N \times 2$, being $N$ the size of the input layer of the CNN. The RFF community attributes the performance drop {\em mainly} to the variability of the radio channel, thus proposing several techniques to mitigate the impact of radio channel impairments on classification accuracy.
%
\\
{\bf Contribution.} In this paper, we provide several contributions. First, we reproduce the \ac{DAT} effect in the same setup of previous works, while we identify and expose another root cause of the performance loss behind the \ac{DAT} effect itself by considering several experiments in different wired and wireless scenarios. Although our analysis confirms that channel impairments affect classification accuracy, we prove that the measurement methodology also has a significant impact on performance. Specifically, we show that the radio's power cycle, i.e., switching on and off both the transmitter and the receiver, significantly affects the classification accuracy when training and testing are performed on datasets collected before and after the power cycle of the radio itself. We verified our assumptions with a wired link between the transmitter and the receiver, thus excluding all the phenomena associated with radio channel variability.
Subsequently, we consider a wireless link running for several days and propose a new methodology to mitigate the \ac{DAT} effect, exploiting the pre-processing of the I-Q samples. Inspired by recent results in the area, we refined the technique of converting the I-Q samples into images, achieving accuracy values that are (on average) twice better than the ones experienced with raw I-Q samples, while halving the variability associated with the reported accuracy. The data used for our analysis are available on request.
\\
{\bf Roadmap.} The paper is organized as follows. Sect.~\ref{sec:related_work} reviews some related work on the robustness of RFF, Sect.~\ref{sec:preliminaries} introduces preliminary concepts, Sect.~\ref{sec:measurement_collection} provides the details of our measurement campaign, Sect.~\ref{sec:the_day_after_tomorrow_effect} provides an in-depth analysis of the \ac{DAT} effect describing the impact of radio power cycle behind it. We also propose a pre-processing technique applying to the I-Q samples to mitigate the \ac{DAT} effect when the power cycle is strictly required. Sect.~\ref{sec:discussion} summarizes our findings and limitations and, finally, Sect.~\ref{sec:conclusion} concludes the paper and outlines future work.

\section{Related Work}
\label{sec:related_work}
\ac{RFF} strategies include a set of techniques to identify and authenticate RF devices by using distinctive patterns in emitted signals~\cite{jagannath2022_arxiv}. Such patterns originate from hardware imperfections in the devices introduced during the manufacturing processes. In the early stage, research on RFF focused primarily on developing custom feature extraction methods using \ac{ML} and \ac{DL} techniques, as shown by~\cite{riyaz2018deep}, \cite{sankhe2019oracle},~\cite{merchant2018deep},~\cite{yu2019robust}, and~\cite{ding2018specific}, to name a few. Although significant progress has been achieved in the development of highly accurate methods for extracting RF features from over-the-air signals, the deployment of RFF systems in the real world faces many challenges. One major limitation of RFF systems based on \ac{DL} algorithms is their sensitivity to wireless channel variability, which can negatively affect their performance. This problem has been reported in several studies~\cite{alshawabka2020}, where the authors found that training a \ac{DL} model on data collected in one day and then testing it on data collected in a different day significantly reduces the classification accuracy. In their experiment, the authors trained three DL models on a dataset of 20 wireless transmitters collected over several days in different environmental settings, including a cable, an anechoic chamber, and in the wild. They showed that the performance of the models was not consistent on different days, indicating that the models were unable to generalize well to new environments or conditions. Throughout this paper, we refer to such phenomenon as the \ac{DAT} effect, causing a dataset shift~\cite{moreno2012unifying}. The findings cited were corroborated by the work of the authors in~\cite{hanna2022wisig}, who also found that changing the receiver used to collect RF signals during training can further degrade the performance of the models. Such findings highlight that the final captured RF fingerprint is a combination of three distinct factors: the transmitter's emitted signal, the channel, and the receiver's hardware. The drop in performance when training and testing on different datasets has also been observed by~\cite{hamdaoui2022}. The authors explained such phenomenon as ``changes in channel conditions", as per Fig.4 in~\cite{hamdaoui2022}. We stress that neither~\cite{hamdaoui2022} nor~\cite{alshawabka2020} provided the details on the power cycle of the considered devices, making the comparison with this work impossible.
The authors in~\cite{elmaghbub2021lora} studied the sensitivity of RFF systems for LoRa networks in a wide range of scenarios, including indoor and outdoor environments, wired and wireless setups, various distances, configurations, hardware receivers, and locations. According to previous research, they found that testing on different days and using different receivers can significantly impact the accuracy of the RFF process. Furthermore, they observed that the variability of protocol configuration and location also has notable effects on the achievable classification accuracy. To address the challenges cited, several mitigation techniques have been proposed in the literature. An approach is to augment the training data by exposing the fingerprinting process to a variety of channel conditions and environments, as demonstrated in~\cite {soltani2020more,al2021deeplora,9966888}. In particular, \cite{shen2021radio} investigated how carrier frequency offset (CFO) affects \ac{RFF} while proposing an ad-hoc classification algorithm to mitigate the phenomenon. Another approach uses the idea of injecting unique impairments into the transmitted signal, such as in ~\cite{sankhe2019oracle,mohanti2020airid,rajendran2020injecting}. Other methods involve using channel modelling and simulations, as in ~\cite{yan2022rrf}, or using digital signal processing techniques with specialized filters, as in \cite{restuccia2019deepradioid,restuccia2021deepfir}. However, none of these works provides an in-depth analysis of the DAT effect; indeed, they do not consider the impact of the radio power cycle on the RFF and they do not provide any specific solution that can deal with a real deployment. In fact, although \ac{RF} fingerprinting is a promising technique for authenticating RF devices at the PHY layer, it still faces many limitations and challenges. In this work, we further advance the analysis of the performance of the \ac{RFF} in real-world scenarios by shedding light on the root causes that affect its performance. Finally, we compare our contribution with the reference literature as per Table~\ref{tab:related_work} in terms of the adopted communication technology, receiver radio hardware, adopted neural network, communication medium, measurement duration, and finally experienced phenomena.

\begin{table*}[]
\caption{Comparison with related work.
}
\label{tab:related_work}
\resizebox{\textwidth}{!}{%
\begin{tabular}{|lllllll|}
\hline
\multicolumn{1}{|l|}{\textbf{Reference}} & \multicolumn{1}{l|}{\textbf{Comm. Tech.}} & \multicolumn{1}{l|}{\textbf{Receiver Radio Tech.}} & \multicolumn{1}{l|}{\textbf{Network}} & \multicolumn{1}{l|}{\textbf{Comm. Medium}} & \multicolumn{1}{l|}{\textbf{Measurement duration}} & \textbf{Observed phenomena}                                                           \\ \hline
\cite{hamdaoui2022}                      & LoRa                                      & USRP B210                                          & Self-designed CNN                     & Wireless                                   & Multiple days                                      & \begin{tabular}[c]{@{}l@{}}Impact of environment and \\ measurement time\end{tabular} \\
\cite{shen2021radio}                     & LoRa                                      & USRP N210                                          & Self-designed MLP/CNN/LSTM            & Cable                                      & Multiple days                                      & Impact of measurement time                                                            \\
\cite{alshawabka2020}                      & WiFi/ADS-B                                & USRP N210 / X310                                   & Self-designed and ResNet50            & Wireless                                   & Multiple days                                      & Impact of measurement time                                                            \\
\cite{al2021deeplora}                      & LoRa                                      & USRP N210                                          & Self-designed LSTM/CNN                & Wireless                                   & Few days                                           & Impact of measurement time                                                            \\
\cite{restuccia2019deepradioid,restuccia2021deepfir}      & WiFi/ADS-B                                & USRP X310                                          & Self-designed CNN                     & Wireless                                   & Multiple days                                      & Impact of channel conditions                                                          \\
\cite{riyaz2018deep}                         & IEEE802.11ac                              & USRP B210                                          & Inspired by AlexNet                   & Wireless                                   & Same day                                           & \begin{tabular}[c]{@{}l@{}}Impact of channel conditions and \\ distance\end{tabular}  \\
\cite{sankhe2019oracle}                  & WiFi                                      & USRPX310                                           & Self-designed CNN                     & Wireless                                   & Same day                                           & Impact of channel conditions                                                          \\
\cite{merchant2018deep}                  & ZigBee                                    & Rohde \& Schwarz FSW67                             & Self-designed CNN                     & Wireless                                   & Same day                                           & Impact of channel conditions                                                          \\
\cite{yu2019robust}                      & ZigBee                                    & USRP N210                                          & Self-designed CNN                     & Wireless                                   & Same day                                           & Impact of channel conditions                                                          \\
\cite{ding2018specific}                  & PHY - QPSK                                & USRP N210 / B210                                   & Self-designed CNN                     & Wireless                                   & Same day                                           & Impact of channel conditions                                                          \\
\cite{shen2021radio} & LoRa & USRP N210 & Self-Designed CNN & Wireless & Multiple days & Impact of CFO \\ \hline
\textbf{Our contribution}                & \textbf{PHY - BPSK}                       & \textbf{USRP X310}                                 & \textbf{ResNet50}                     & \textbf{Wireless and Cable}                & \textbf{Multiple days}                             & \textbf{\begin{tabular}[c]{@{}l@{}}Impact of channel conditions and \\ power cycle\end{tabular}}                                                          \\ \hline
\end{tabular}%
}
\end{table*}

\section{Preliminaries}
\label{sec:preliminaries}
In this section, we introduce the hardware and software setup adopted in this work and the required background concepts related to \ac{RF} communication and \ac{DL} for radio fingerprinting. 

\begin{figure}
    \centering
    \includegraphics[width=\columnwidth]{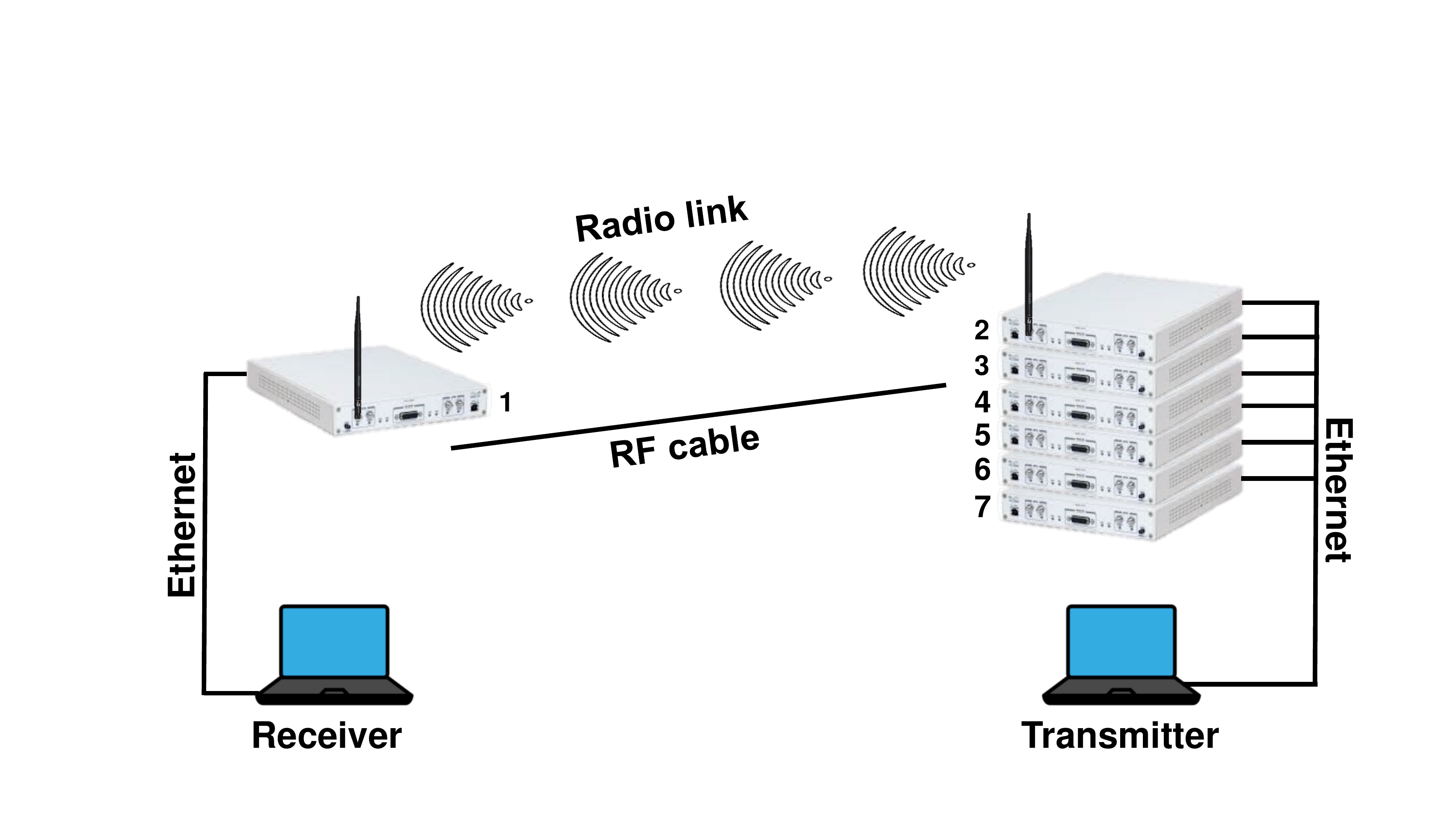}
    \caption{Hardware setup: our measurement setup consists of 1 receiver and 6 transmitters communicating via either a wireless or a wired link, depending on the scenario.}
    \label{fig:hw_setup}
\end{figure}

{\bf Hardware Set-up.} The hardware considered in this paper includes seven (7) \acp{SDR} USRP X310~\cite{ettus}, featuring the UBX160 daughterboard and the VERT900 antenna~\cite{ubx}. The general setup is depicted in Fig.~\ref{fig:hw_setup}, showing that the radios are connected to two laptops HP EliteBook I7, featuring 32GB of RAM. All the considered scenarios involve radio $1$ (on the left side in Fig.~\ref{fig:hw_setup}) as the receiver, while we consider the other ones as the transmitters (only one active for each experiment). Depending on the specific scenario,  we connected the transmitter and the receiver via either an \ac{RF} (wireless) or a wired link. In the former, we considered a transmitting power of 35mW and a normalized receiver gain of 1, while when using a wired link we set the transmission power to 1 mW and the normalized receiver gain to 0.8, where the normalized receiver gain is defined according to the logic in the USRP Source block provided by GNURadio (see software setup below). Finally, for the wired link, we considered a coaxial cable type RG58A/U. We do note that other work in the RFF research area might consider a larger number of transmitter radios. However, our findings confirm their performance, while paving the way for future research in the area.

{\bf Software Setup.} We adopted GNURadio v3.8 for the measurements, defining a transmission chain with the following blocks:
\begin{itemize}
    \item {\em File source.} We generated a message including a string of 256 bytes with incremental value, i.e., $[0, \ldots, 255]$. Note that specifying a valid message is necessary to draw any conclusions on the effectiveness of the TX-RX chain, as well as on the bit-error rate experienced by the communication link.
    \item {\em Constellation modulator.} We configured the modulator according to the \ac{BPSK} modulation.
    \item {\em UHD: USRP Sink.} We set the transmission frequency to $900$~MHz (carrier), with a sample rate of 1M samples per second, and finally, the transmission power of either 35mW or 1mW for the wireless and the wired link, respectively.
\end{itemize}
We configured the receiver according to the following chain:
\begin{itemize}
    \item {\em UHD: USRP Source.} We set the receiving frequency at $900$~MHz (carrier), with a sample rate of 1M samples per second and a normalized receiver gain of 1 or 0.8 depending on the adopted link, being wireless or wired, respectively. Note that when receiving complex I-Q values, the modified Nyquist theorem requires that the sample rate at the receiver should be at least equal (or higher) to the sample rate at the transmitter~\cite{sampling_same}. This is due to the fact that at each sampling instant, we acquire two samples (I and Q). Note that this is the standard configuration to receive and demodulate a 1M bandwidth signal transmitted at the frequency of 900MHz.
    \item{\em AGC.} We included the Adaptive Gain Control block to mitigate channel fluctuations.
    \item{\em Symbol Sync.} We included a symbol synchronizer to receive and decode digital signals based on the \ac{mle} criterion.
    \item{\em Costas Loop.} We adopted the Costas loop to mitigate the phase offset and residual frequency offsets.
    \item{\em File Sink.} The output of the receiving chain is stored inside a file, for follow-up analysis.
\end{itemize}

\begin{figure}
    \centering
    \includegraphics[width=\columnwidth]{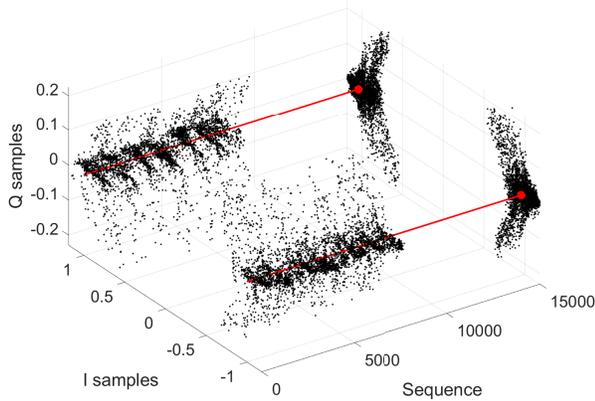}
    \caption{I-Q samples: solid red lines represent the theoretical position of the I-Q samples for a BPSK modulation scheme, i.e., [-1, 0] and [1, 0], while the dispersion of their values is mainly caused by the transducers' imperfections---these samples are coming from a wired measurement---thus representing the fingerprint of the radio.}
    \label{fig:iq_samples}
\end{figure}

{\bf I-Q Samples.} Given the carrier frequency $f_0$, which in our case is $900$~MHz, a digital modulation scheme can be described through Eq.~\ref{eq:modulation}~\cite{lathi}.
\begin{equation}
    x(t) = I \cos{(2\pi f_0 t)} + Q \sin{(2\pi f_0 t)},
    \label{eq:modulation}
\end{equation}
where $x(t)$ is the transmitted modulated signal, $I$ is the \emph{in-phase} component, while $Q$ is the \emph{quadrature} component. For the specific modulation scheme considered in this work, that is, \ac{BPSK}, the quadrature component always has a null value ($Q = 0$), while the in-phase component is used to translate the value of the bit $b$, that is, $I = - 1$ and $Q = 0$ when $b = 0$, and $I = + 1$ and $Q = 0$ when $b = 1$, or vice versa, as shown by Eq.~\ref{eq:bpsk}.
\begin{align}
    x(t) & =   
    \begin{cases}
        +1 \cos{(2\pi f_0 t)}, & \text{if } $b == 1$ \\
        -1 \cos{(2\pi f_0 t)}, & \text{if } $b == 0$ \nonumber
    \end{cases}\\
        & = \cos{(2\pi f_0 t + \phi)},
    \label{eq:bpsk}
\end{align}
where $\phi$ takes on the value of either 0 or $\pi$ as a function of the value of the bit. Since $I$ and $Q$ constitute an alternative way of representing the magnitude and phase of the modulated signal $x(t)$, it is natural to consider the components $I$ and $Q$ as the real and imaginary parts of a complex number, respectively. In particular, for the BPSK, there is no imaginary part ($Q=0$), while we only consider the real component, that is, $I = \pm 1$. Given a sequence of bits, the transmitter implements Eq.~\ref{eq:bpsk} to translate bits into I-Q samples, while the receiver takes on the challenge of reversing each IQ sample to the original value of the bit. As a toy example, we consider Fig.~\ref{fig:iq_samples}, which represents a sequence of I-Q samples over time. The measurement consists of 8000 I-Q samples (black dots) collected by using a wired link between the transmitter and the receiver, while the red lines and the associated red dots (at the end of the lines) represent the theoretical (ideal) position of the I-Q samples, that is, [1, 0] and [-1, 0]. Due to radio imperfections, the I-Q samples spread over the I-Q plane with a specific pattern peculiar to the adopted combination of transmitter and receiver radios. Finally, we report the projection of the I-Q samples at the bottom of the figure to provide a more concrete representation of the actual spreading of the I-Q samples, i.e. the \emph{radio fingerprint}.

{\bf Deep Learning.} We denote {\em radio fingerprinting} as the task of identifying unique features at the physical-layer (I-Q samples) that can be used to discriminate a radio transmitter. A radio fingerprint is a unique pattern in the I-Q samples, like the one depicted in Fig.~\ref{fig:iq_samples}. Many researchers have taken on the challenge of identifying a robust fingerprint and developing a methodology to use it efficiently and effectively. State-of-the-art solutions involve a two-stage process: (i) {\em training} a neural network model with chunks of I-Q samples, and (ii) {\em testing} a sequence (of I-Q samples) from the wild to identify the actual transmitter. Among the several neural networks proposed in the literature, the family of \ac{RNN} emerged as a good trade-off between training speed and classification performance. In particular, {\em resnet50} is the one adopted in this work, in line with other contributions which highlighted the temporal variation of the fingeprint \cite{alshawabka2020}, \cite{hamdaoui2022}, \cite{oligeri2022}, while many other contributions adopt similar networks, changing a subset of its layers. The choice of {\em resnet50} (as also discussed in~\cite{oligeri2022}) is based on an empirical verification of its performance while considering different input configurations.
Specifically, we consider the {\em resnet50} implemented in MatLab R2022b\textsuperscript{\tiny\textregistered}, constituted by 50 layers and pre-trained on the ImageNet database. The neural network, as originally designed, cannot be used for the training and classification of the I-Q samples, and researchers developed different ways to modify it to fit the radio fingerprinting problem. Indeed, {\em resnet50} is designed to classify images from the ImageNet dataset, thus requiring some modifications at both the input and output layers. The input layer is usually adapted to fit raw I-Q samples (and not images), while the output layer is changed to fit the number of transmitters, being different from the 1000 classes of ImageNet. Moreover, the network itself is already trained to classify images such as dogs, cats, flowers, etc.---these being different from the input adopted for the \ac{RF} fingerprinting; therefore, the network requires a partial re-train to expose the model to the new input.
\\
In detail, the output layers ({\em fullyConnectedLayer} and {\em classificationLayer}) should be re-adapted to take into account the number of classes in the radio fingerprinting problem, i.e., the number of radio transmitters to identify. Regarding the input layers, two main configurations have been considered: (i) input of interleaved raw I-Q samples consisting of a vector of dimensions either $N \times 1 \times 1$, as in~\cite{alshawabka2020}, or $N \times 2 \times 1$, as in~\cite{hamdaoui2022}, and (ii) images where the input is a matrix of size $224 \times 224 \times 3$, where $224 \times 224$ is the size of the images in the ImageNet dataset, as in~\cite{oligeri2022}.

\begin{figure}
    \centering
    \includegraphics[width=\columnwidth]{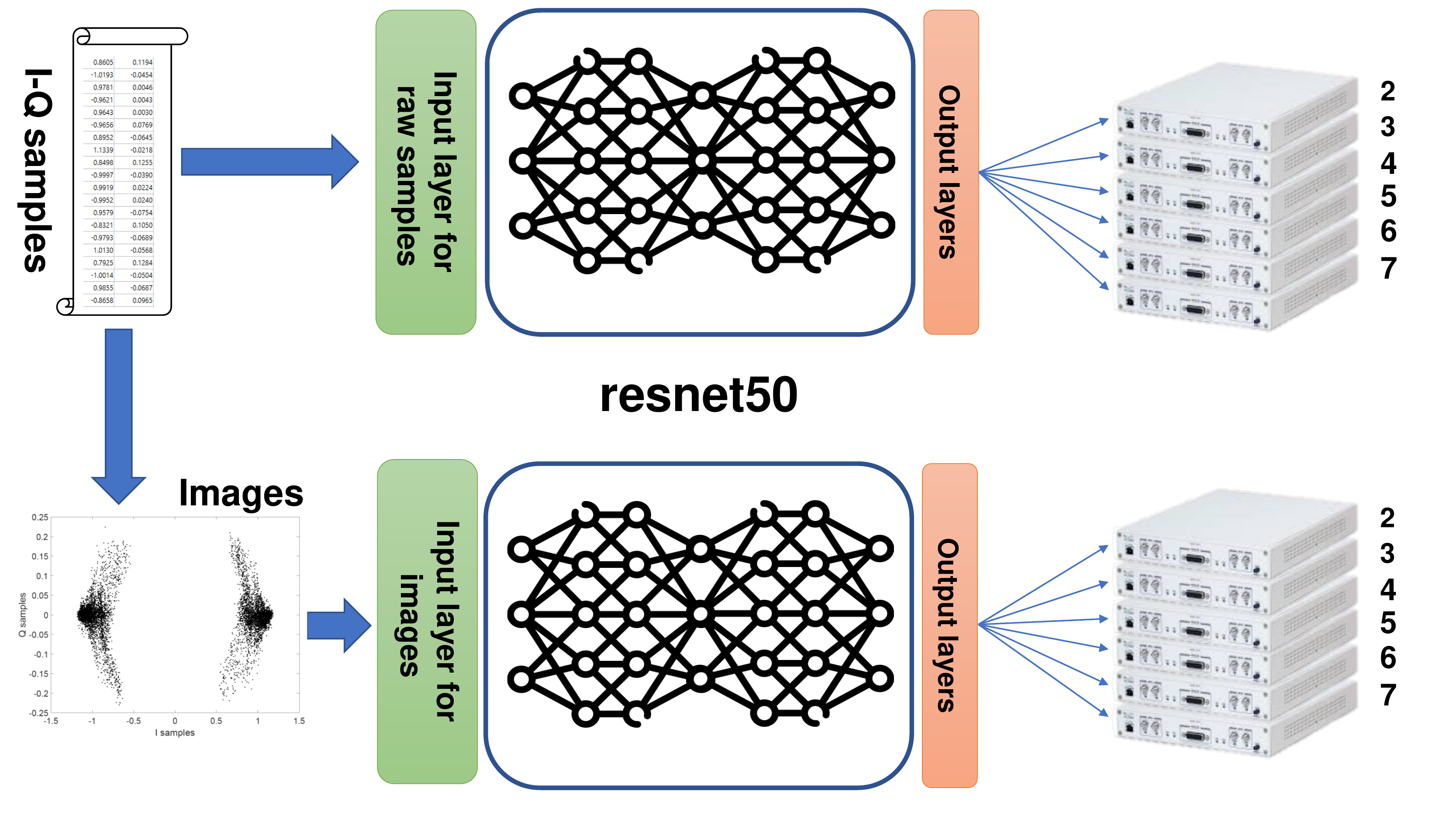}
    \caption{The RFF process: a neural network ({\em resnet50}) is re-adapted to fit either raw I-Q samples or images (input layers) and to identify the correct number of radios (output layers).}
    \label{fig:classification_process}
\end{figure}

In the remainder of this paper, we will consider both the approaches of raw I-Q samples and images, to compare the performance and highlight their limitations. Figure~\ref{fig:classification_process} shows how we deployed {\em resnet50} to identify the transmitter given the I-Q samples collected by the receiver, considering the pre-processing approaches discussed above.
\section{Measurement collection}
\label{sec:measurement_collection}
We define {\em measurement} a continuous flow of I-Q samples between the transmitter and the receiver. As will be clarified in the following, we consider two types of measurements: (i) short measurements lasting 300 seconds (5 minutes) and (ii) long-term measurements lasting 4 days. In the remainder of this work, we refer to I-Q samples as complex-valued data; thus, each radio sample (bit) is constituted by one I and one Q sample. Given the availability of 6 transmitters, we define {\em run} of measurements as the sequence of 6 consecutive measurements where the receiver is the same while the transmitter changes every time among all the available ones. Finally, we define {\em dataset} as a sequence of several runs of measurements. All the datasets characterized by a wireless link share the same scenario, that is, an indoor environment where the transmitter is located 10 meters from the receiver, without the line-of-sight (nLoS). During our measurement campaign, we collected three datasets of measurements, listed below and summarized in Tab.~\ref{tab:measurements}. All the measurements can be requested by email to the authors of the paper.
\begin{enumerate}
    \item {\em Dataset 1 (DS1).} This dataset is constituted of 78 measurements, organized in 13 runs, collected by using a wired link and power-cycling the radios (both the transmitter and the receiver) after each measurement. We kept the same receiver for all measurements while we changed the transmitter (using 6 different radios, in total).
    \item {\em Dataset 2 (DS2).} This dataset is constituted by 6 non-stop measurements collected by using a wired link for 3 days for each measurement ($6 \times 3 = 18$ days, 265B+ samples). All the measurements share the same receiver, while we changed the transmitters (using 6 different radios in total, as for the previous dataset). Given the duration of the measurements, we had to decrease the sample rate to 256K samples per second.   
    \item {\em Dataset 3 (DS3).} This dataset is constituted by 72 measurements, collected using the wireless radio link (using the carrier frequency $f=900$~Mhz) over 4 days with a distance between the transmitter and the receiver of 10 meters. The dataset has been collected on the ground floor of a villa where the transmitter has been set up in the living room and the receivers in the kitchen (no line of sight), with 2 people randomly crossing the measurement set-up.
    For each day, we collected 3 runs of measurements (6 measures each) early in the morning, noon, and late in the evening, power-cycling the devices in-between the measurements. To be consistent with the other measurements, we considered the same radio as the receiver, while we swapped the transmitters among the 6 available radios.
\end{enumerate}
\begin{table}
\caption{Measurements description: We collected three datasets over wired and wireless links, lasting multiple days.}
\label{tab:measurements}
\footnotesize
\begin{tabular}{l|l|c|c|c|c|}
\cline{2-6}
    & {\bf Link} & \begin{tabular}[c]{@{}l@{}}{\bf Sample}\\ {\bf Rate [Msps]}\end{tabular} & \begin{tabular}[c]{@{}l@{}}{\bf Duration}\\ {{\bf [Days]}}\end{tabular} & {\bf Runs} & \begin{tabular}[c]{@{}l@{}}{\bf Samples}\\ {\bf per}\\ {\bf Measurement}\end{tabular} \\ \hline
\multicolumn{1}{|l|}{\em DS1} & Wired & 1     &  3 & 13 & 144M+  \\ \hline
\multicolumn{1}{|l|}{\em DS2} & Wired & 0.256 & 18 & 1  & 33B+  \\ \hline
\multicolumn{1}{|l|}{\em DS3} & Radio & 1     &  4 & 12 & 144M+  \\ \hline
\end{tabular}
\end{table}

\textcolor{black}{Note that we consider multiple transmitters, but only a single receiver. This is done on purpose to replicate the traditional setup adopted in the literature for \ac{RF} fingerprinting. Analyzing the impact of a different receiver on the performance of RFF solutions is out of the scope of this contribution, as well as part of our future work.}
\section{The Day-After-Tomorrow effect}
\label{sec:the_day_after_tomorrow_effect}
In this section, we first summarize the methodology followed in our investigation (Sect.~\ref{sec:methodology}), and subsequently, we provide an in-depth analysis of the \ac{DAT} effect (Sect.~\ref{sec:DAT_in_depth}). Moreover, in Sect.~\ref{sec:DAT_mitigation}, we introduce a mitigation technique for the \ac{DAT} effect leveraging the pre-processing of I-Q samples into images and, finally, in Sect.~\ref{sec:radio_link}, we deploy our solution to a real wireless radio link.
In the following, we use the qualitative terms {\em low} and {\em high} associated with the accuracy of the classifier just for the sake of simplicity, as well as to introduce the problem discussed in our manuscript. A quantitative analysis of the phenomena involving real measurements, taken with both wireless and wired links, will be presented in the subsequent sections to support our claims.

\subsection{Methodology}
\label{sec:methodology}
\begin{figure}
    \centering
    \includegraphics[width=\columnwidth]{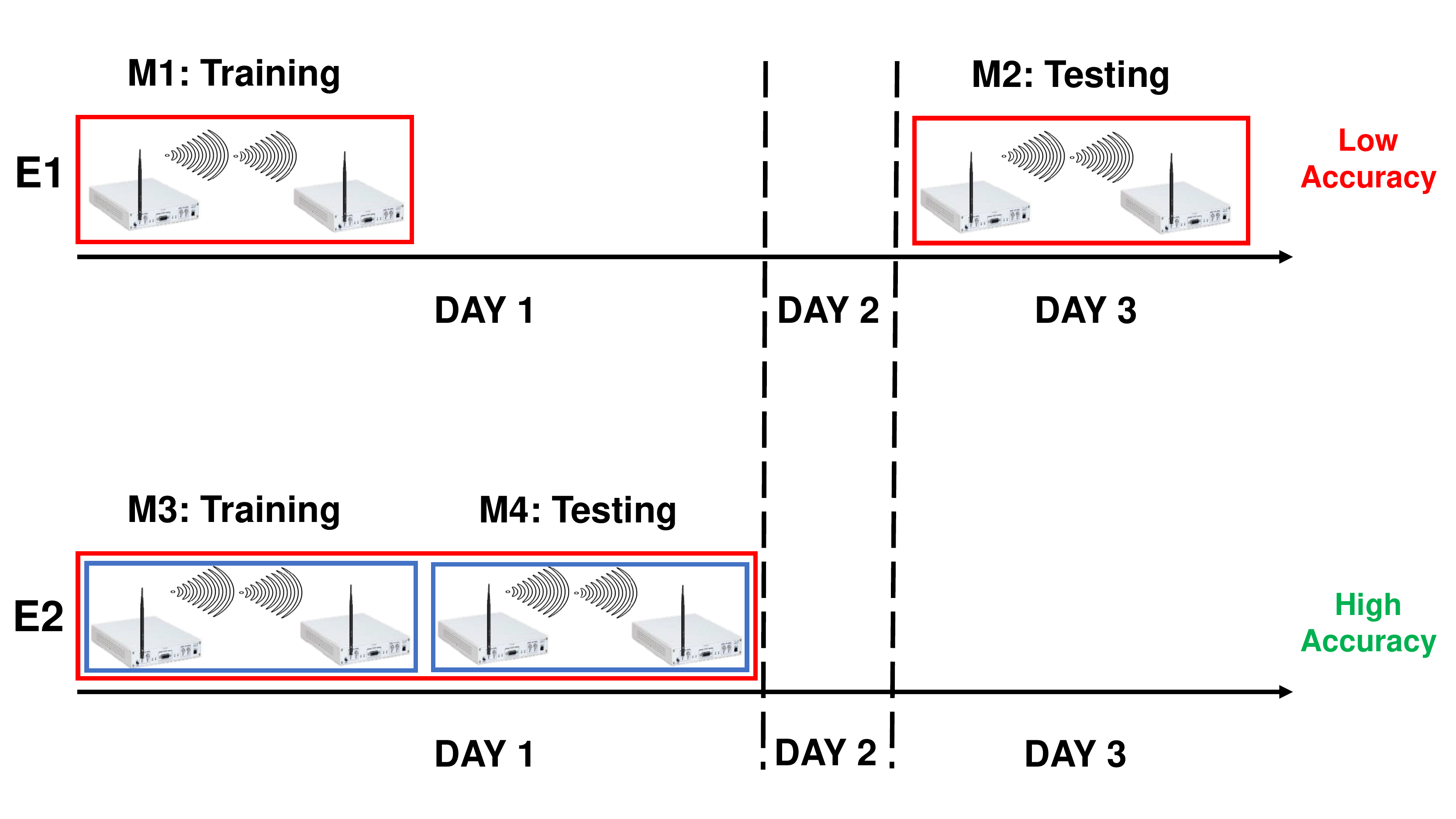}
    \caption{The {\em Day-After-Tomorrow} effect is constituted by two experiments (E1 and E2): training a model on a measurement (M1) taken one day (DAY 1) and then testing on a measurement (M2) taken on another day (DAY 3) gives low accuracy performance. Conversely, training and testing on chunks (M3 and M4) taken from the same measurement (DAY 1) gives high accuracy performance.}
    \label{fig:DAT_effect}
\end{figure}

\begin{figure}
    \centering
    \includegraphics[width=\columnwidth]{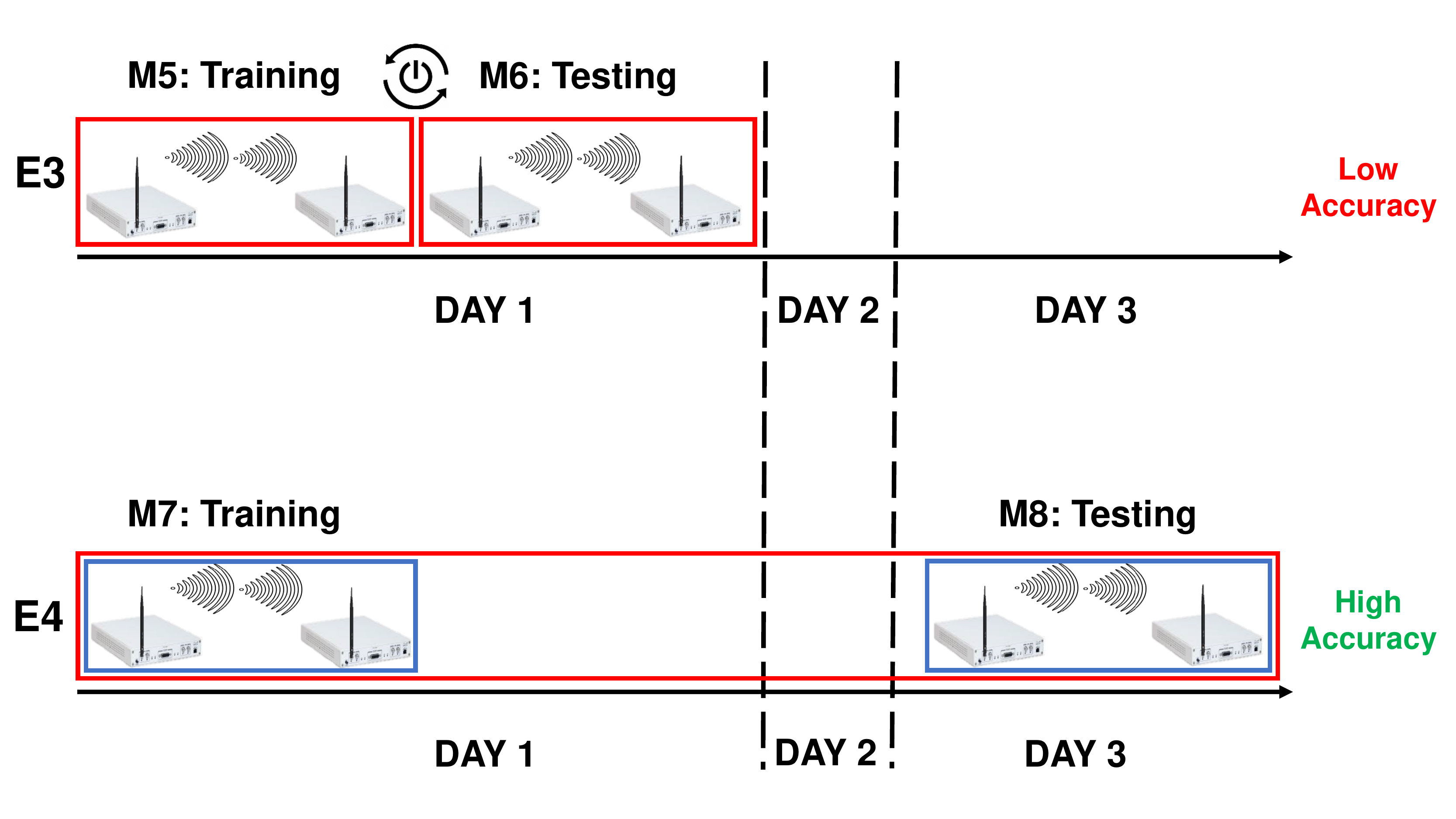}
    \caption{The {\em power cycle} problem: consecutive measurements (M5 and M6) experiencing similar channel conditions achieve low classification accuracy if radios are power cycled in-between the measurements. Conversely, a long-lasting measurement (E4) experiencing very different channel conditions achieves high accuracy if the two chunks (M7 and M8) are from the same measurement, i.e, they are not separated by a power cycle.}
    \label{fig:power_cycle}
\end{figure}

This section discusses the \emph{Day-After-Tomorrow} effect, i.e., a common problem affecting all the approaches dealing with RFF presented in the literature and mentioned explicitly in some of them, e.g.,~\cite{alshawabka2020} and \cite{hamdaoui2022}. We describe the phenomenon through real measurements, whose logic is depicted in Fig.~\ref{fig:DAT_effect} using two experiments (E1 and E2). We highlight that experiments E1 and E2 depict a radio link, but, in the following, we might consider either a wireless or a wired link, depending on the objectives of the experiments. For both experiments (E1 and E2), two measurements (red boxes) are collected. During the first experiment (E1), the measurements (M1 and M2) are taken on two different days (Day 1 and Day 3). In contrast, during the second experiment (E2), only one measurement is taken (one red box) and then split into two chunks (blue boxes), M3 and M4 respectively, taken on Day 1 without interrupting the measurement process. For both experiments, we use one measurement to train a neural network model, i.e., M1 in E1 and M3 in E2, while we consider the second measurement for testing, i.e., M2 in E1 and M4 in E2. 
We refer to the {\em Day-After-Tomorrow} (\ac{DAT}) effect as the phenomenon where the first experiment E1 is characterized by low classification accuracy compared to the second experiment E2, which experiences a high classification accuracy. This is a well-known effect in the literature, reported by recent authoritative contributions such as~\cite{alshawabka2020} and~\cite{hamdaoui2022}, to name a few. Such works explain the \acs{DAT} effect by referring to the unpredictability of the wireless radio channel and its associated phenomena, e.g., multipath and fading. We highlight that the \acs{DAT} effect significantly hinders the deployment of radio fingerprinting techniques in any real-world scenarios since the model trained at a specific time cannot be effectively re-used in the future without reporting significant performance loss. 
While we acknowledge the impact of the radio channel in the fingerprinting process, in the following, we formulate a different hypothesis rooted in real experiments involving both wireless and wired links. Our intuition is that the loss of performance described above is not caused by the channel variations only---although the channel might play an important role---but by another phenomenon, i.e., the {\em power cycle} of the radios in-between the measurements used for the training and the testing tasks. 
\begin{definition}
    We define \emph{power cycle} as the process involving the software (re-)initialization of the radio. This takes place by applying the power-off/power-on of the radio.
\end{definition}
Figure~\ref{fig:power_cycle} summarizes the experiments we conducted to expose the phenomenon described above. Although the experiments depict a radio link, in the following, we might consider either a wireless or a wired link depending on the objectives of the experiments. Specifically, we consider two additional experiments (E3 and E4): during E3, two measurements (M5 and M6) are collected by power-cycling the transmitter and the receiver in-between the measurements, i.e., switching off and on the radios. We stress that M5 and M6 are taken one immediately after the other, i.e., very close in time, by only switching off and on the radios (power-cycling). We use M5 for training and M6 for testing. We observed a low classification accuracy for both the wireless and the wired scenario. Finally, we consider another experiment (E4) constituted by a long measurement (spanning 3 consecutive days) where no power cycle is performed in between the measurements. From this experiment, we extract two chunks, i.e., M7 and M8. When considering M7 for training and M8 for testing, the resulting classification accuracy is high.
In the following sections, we also show that it is possible to mitigate this phenomenon and significantly increase the classification accuracy for measurements separated by a power cycle (E3). This is achieved by pre-processing the I-Q samples and converting them into images, adopting these later ones as the input to the CNN.
\subsection{\acs{DAT}: In-depth analysis}
\label{sec:DAT_in_depth}
\begin{figure*}
\centering
\begin{minipage}{.13\textwidth}
  \centering
  \includegraphics[width=\linewidth]{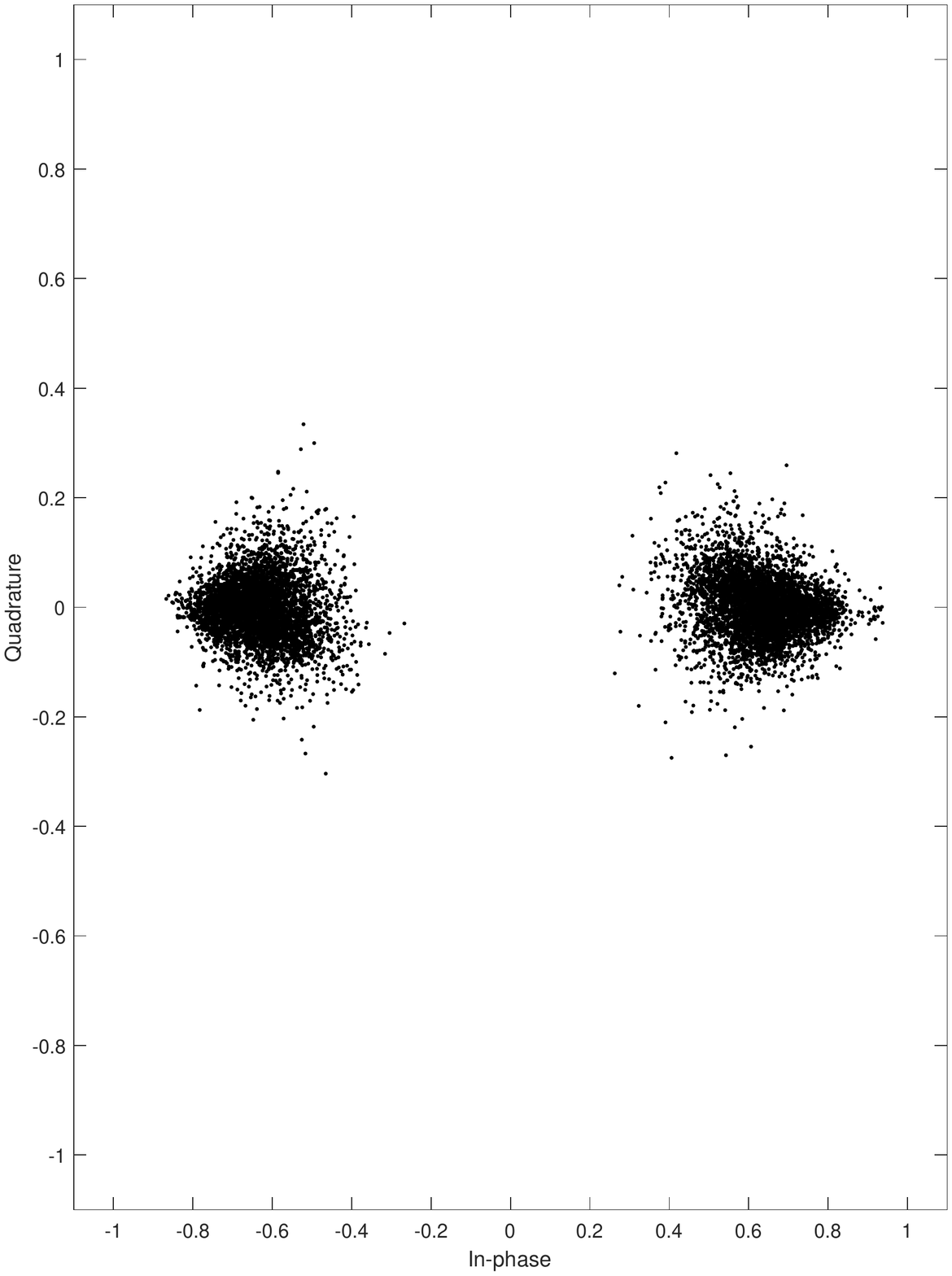}
    \caption*{(a)}{t = 0}
\end{minipage}
\begin{minipage}{.13\textwidth}
  \centering
  \includegraphics[width=\linewidth]{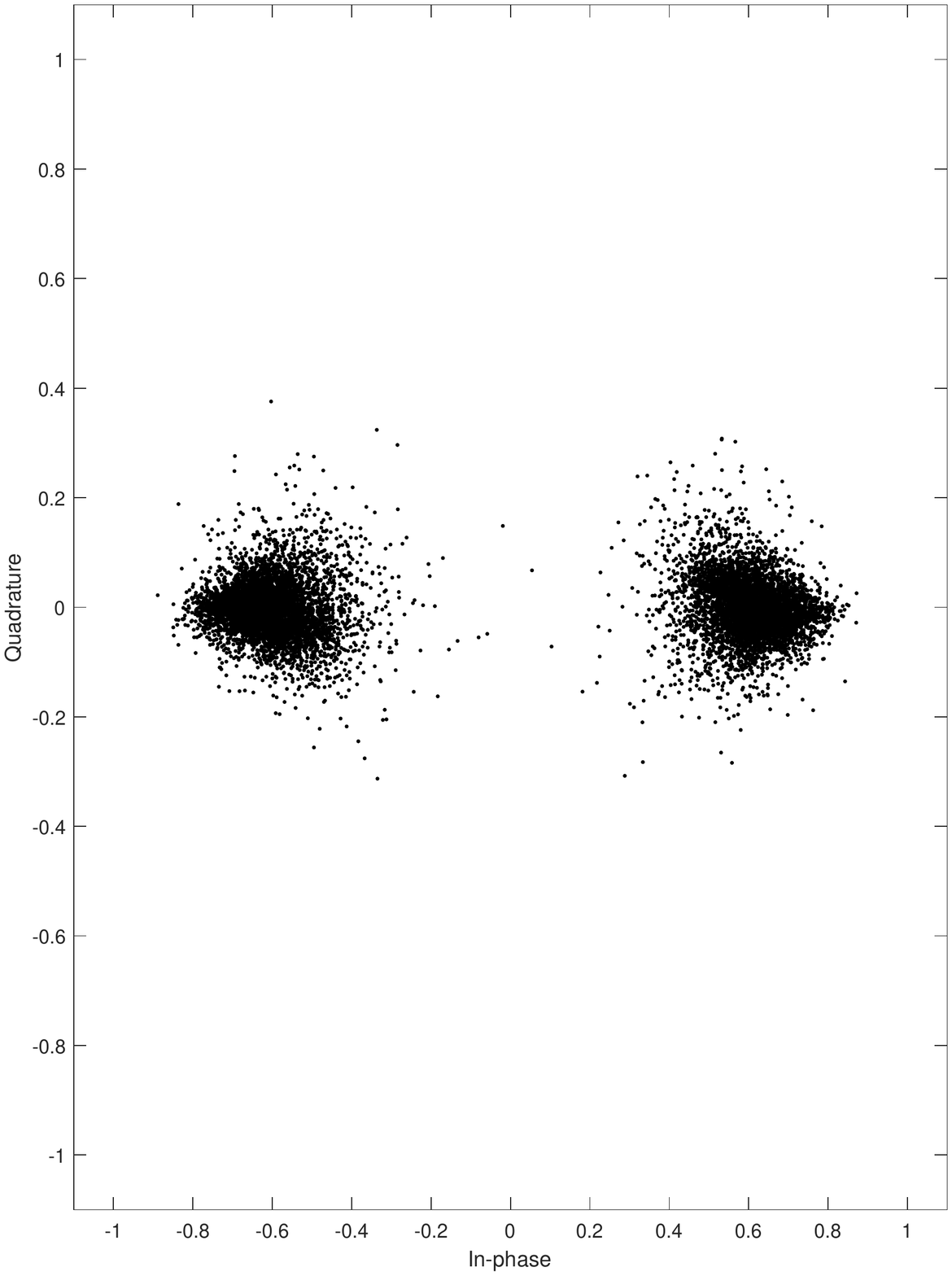}
    \caption*{(b)}{t = 1}
\end{minipage}
\begin{minipage}{.13\textwidth}
  \centering
  \includegraphics[width=\linewidth]{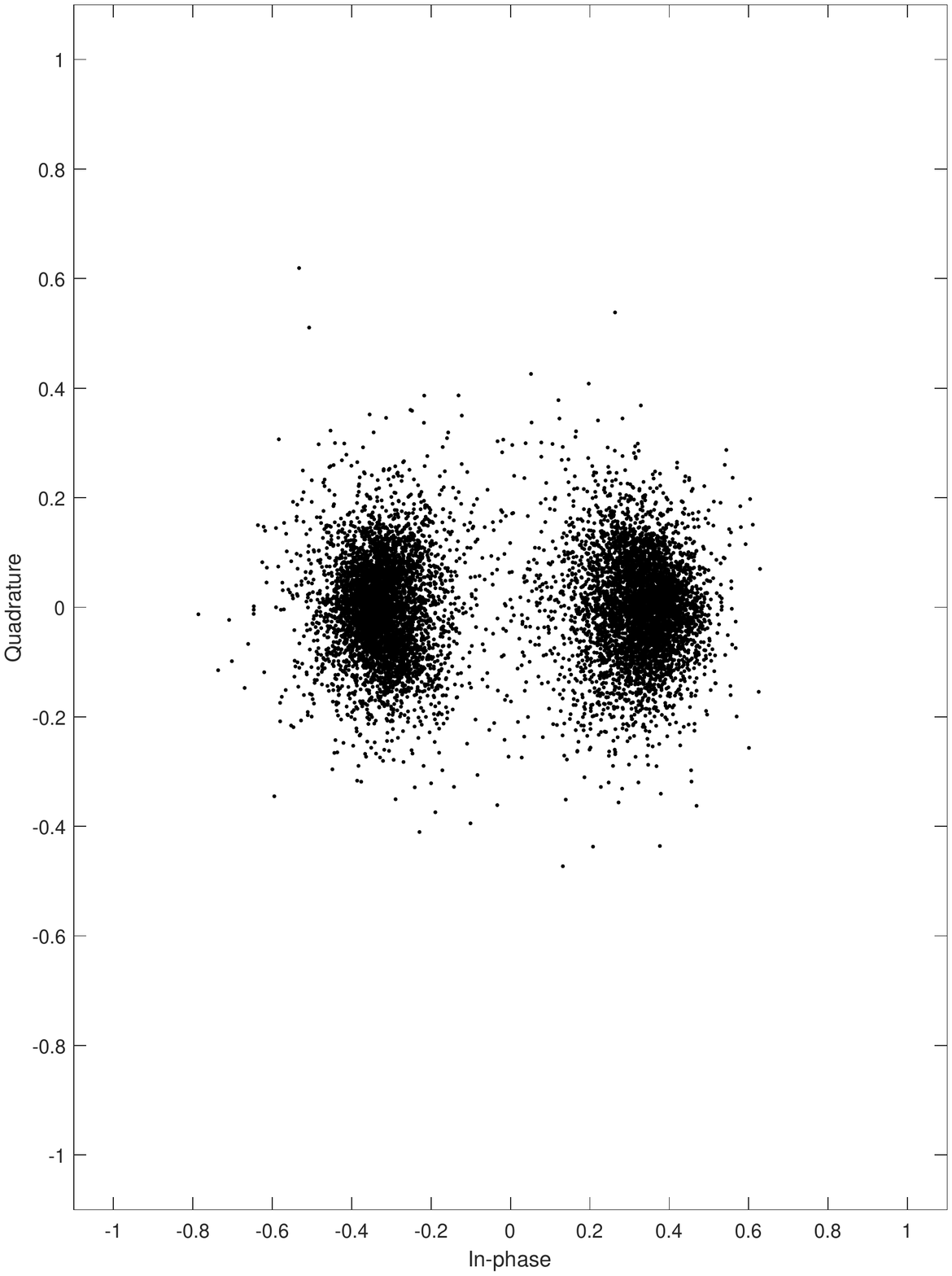}
    \caption*{(c)}{t = 2}  
\end{minipage}
\begin{minipage}{.13\textwidth}
  \centering
  \includegraphics[width=\linewidth]{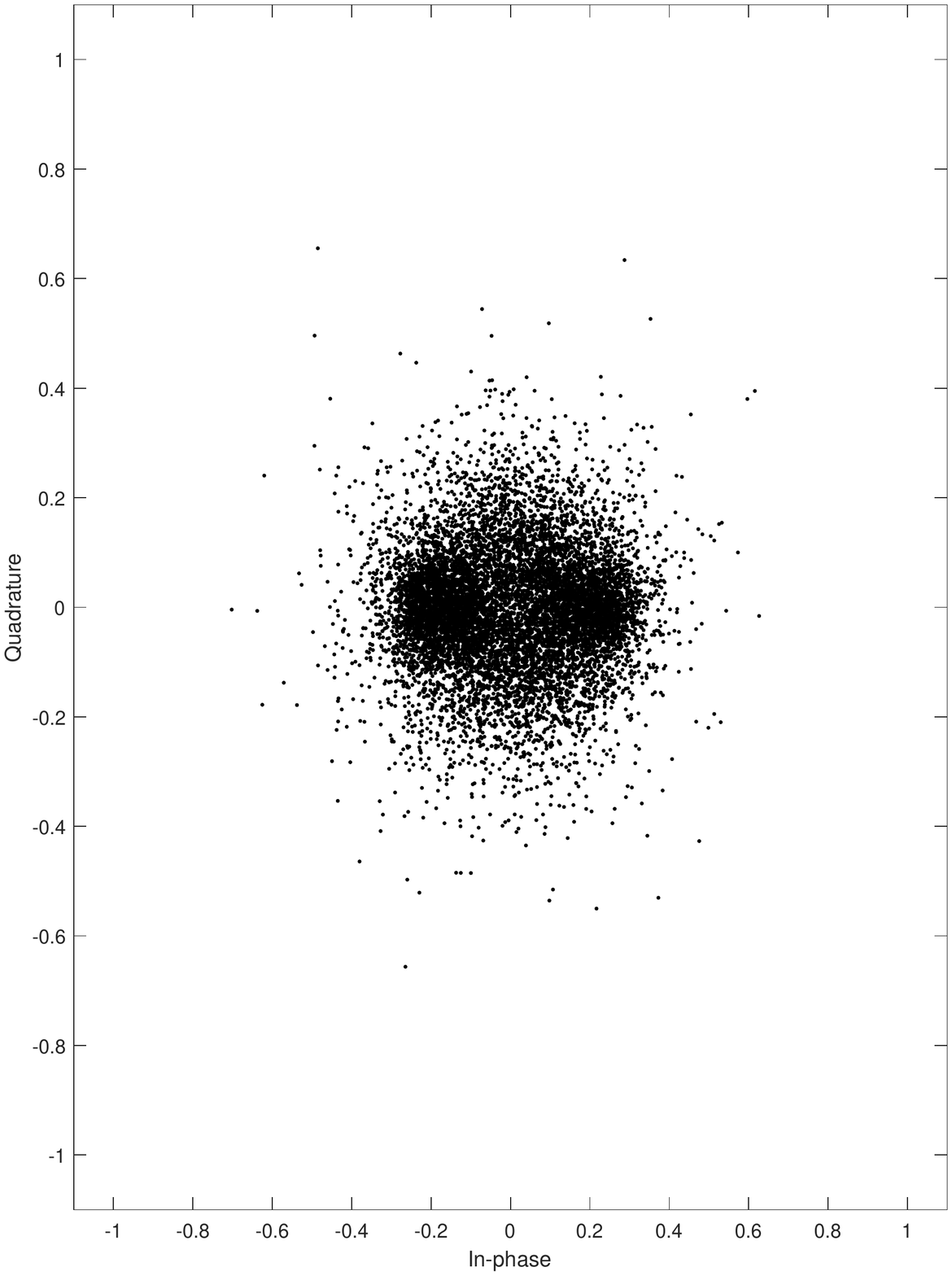}
    \caption*{(d)}{t = 3}
\end{minipage}
\begin{minipage}{.13\textwidth}
  \centering
  \includegraphics[width=\linewidth]{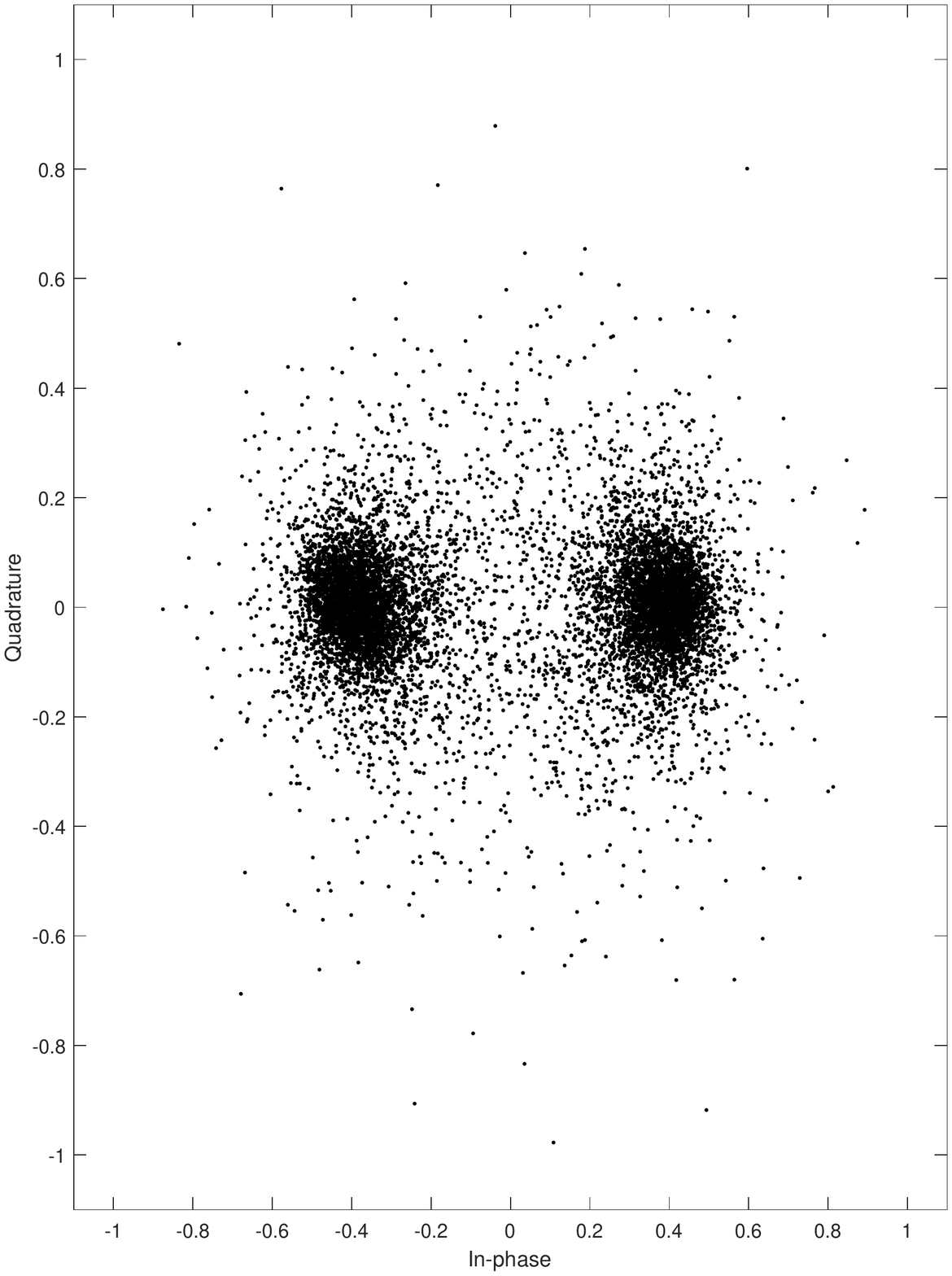}
    \caption*{(e)}{t = 4}
\end{minipage}
\begin{minipage}{.13\textwidth}
  \centering
  \includegraphics[width=\linewidth]{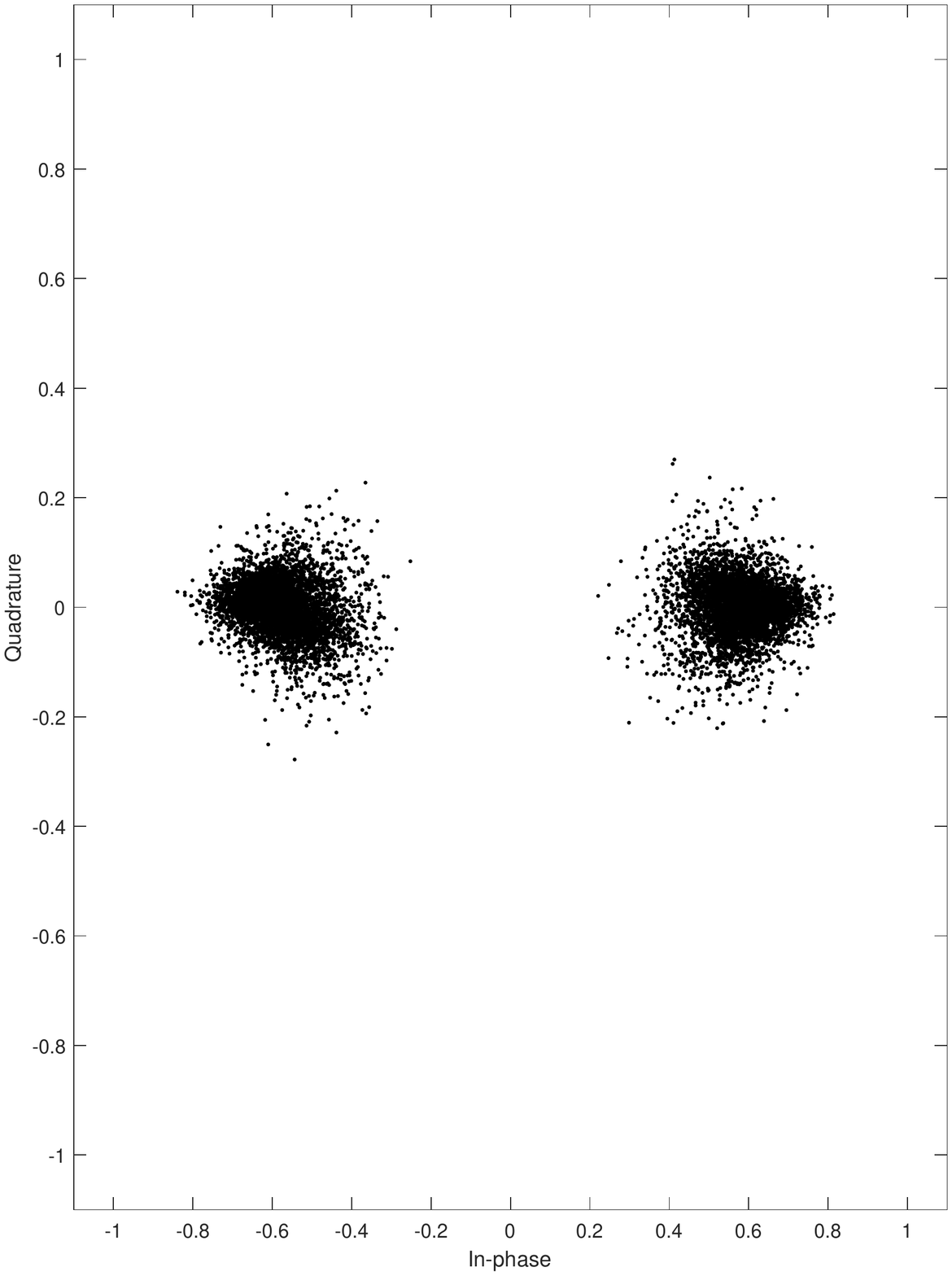}
    \caption*{(f)}{t = 5}  
\end{minipage}
\begin{minipage}{.13\textwidth}
  \centering
  \includegraphics[width=\linewidth]{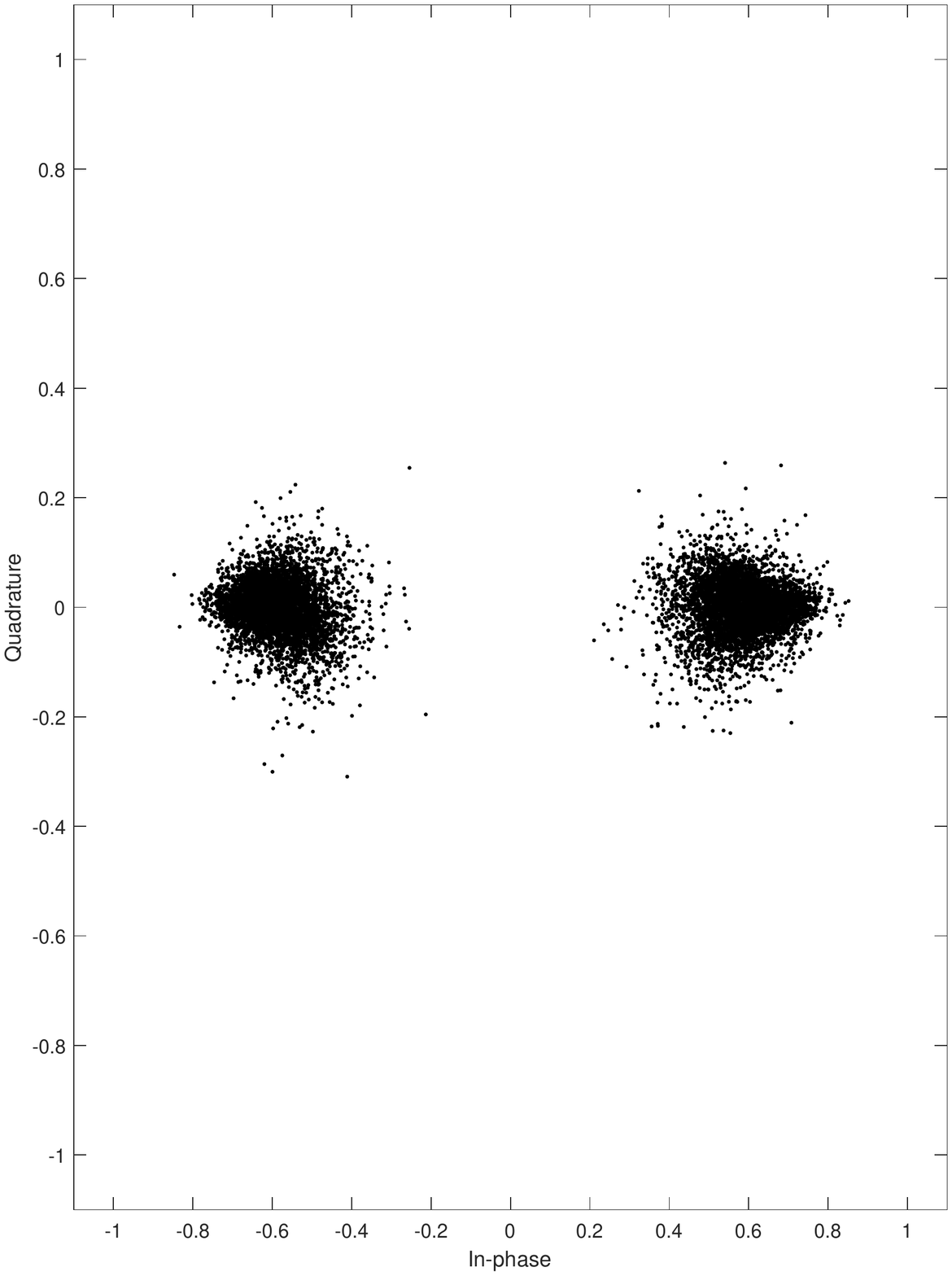}
    \caption*{(g)}{t = 6}
\end{minipage}
\caption{The effect of multi-path to the transmitter-receiver link across seven time windows. Compared to an ideal (static) scenario (Figs. (a), (b), (f), and (g)), the shape of the I-Q samples at the receiver (modulated through the BPSK scheme), is significantly affected (see Figs. (c), (d), and (e)) when an event is affecting the scene, e.g., moving objects.}
\label{fig:multipath}
\end{figure*}

The vast majority of the literature explains the \acs{DAT} effect with the unpredictability of the radio channel, considering the changes that affect the environment surrounding the transmitter and receiver. 
As an example, we consider Fig.~\ref{fig:multipath}, consisting of 7 sub-figures ($100,000$ I-Q samples each) taken from subsequent chunks in an actual radio measurement. Fig.~\ref{fig:multipath}(a) and (g) represent the steady state, before and after the perturbation event, that is, a person walking close to the radio link. It is worth noting that the I-Q samples, and therefore the transmitter's fingerprint, are strongly affected by the considered event. This is evident when looking at the figures from Fig.~\ref{fig:multipath}(c) to (e), and in particular, Fig.~\ref{fig:multipath}(d), showing a set of I-Q samples completely different from the ones at steady state, i.e., Fig.~\ref{fig:multipath}(a) and (g). Phenomena such as the one in Fig.~\ref{fig:multipath}(d) completely change the I-Q displacement at the receiver, making the retrieval of the fingerprint (almost) impossible.
%
To discuss in more depth the \acs{DAT} effect and its causes, we consider a wired link between radios. We consider this scenario as it excludes any dependencies on the multipath and other effects due to RF propagation while keeping the noise due to the communication channel at a minimum. We will extend our results to the RF link in Sect.~\ref{sec:radio_link}.

\begin{figure*}
\centering
\begin{minipage}{.33\textwidth}
  \centering
  \includegraphics[width=\linewidth]{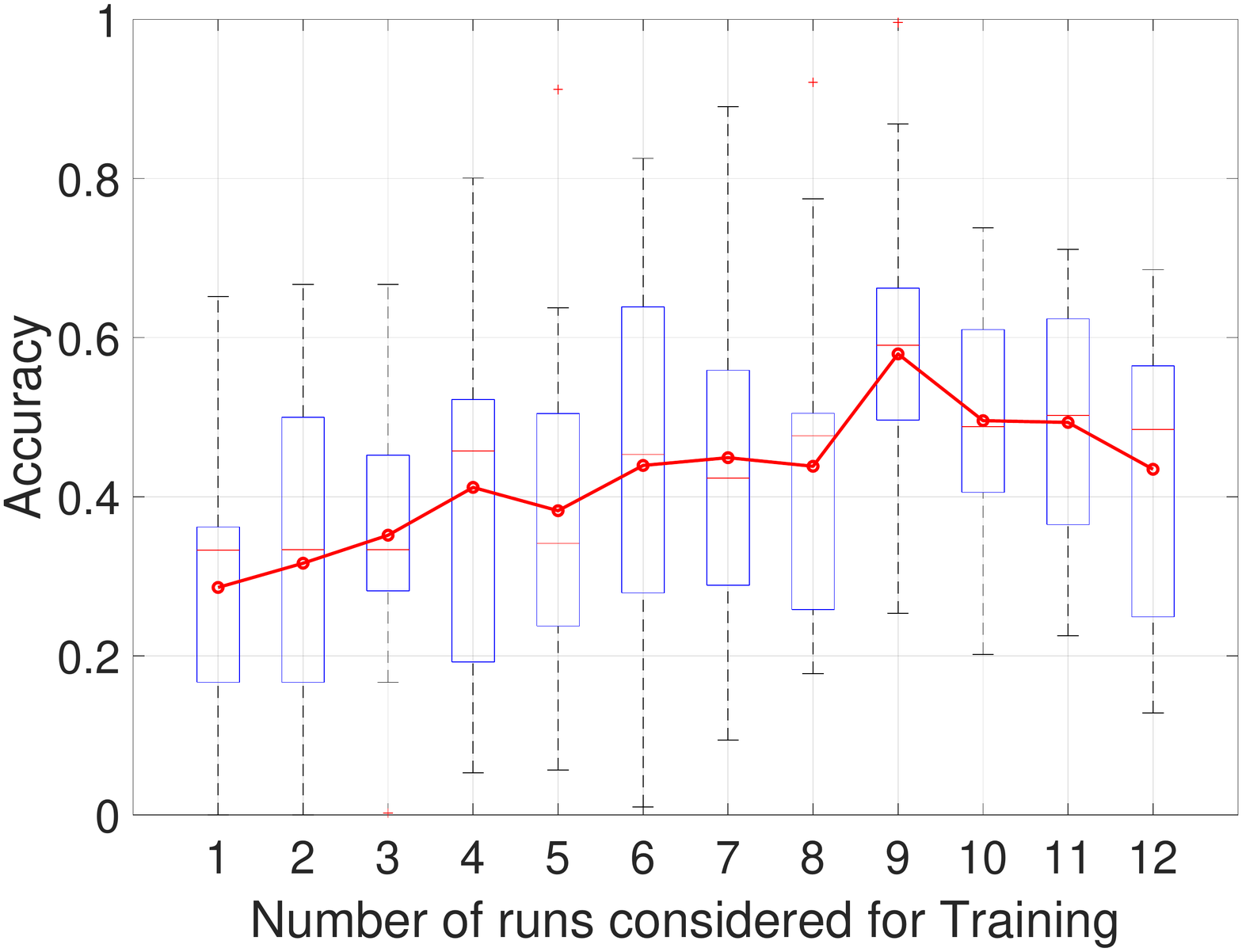}
  \caption*{(a)}
\end{minipage}%
\begin{minipage}{.33\textwidth}
  \centering
  \includegraphics[width=\linewidth]{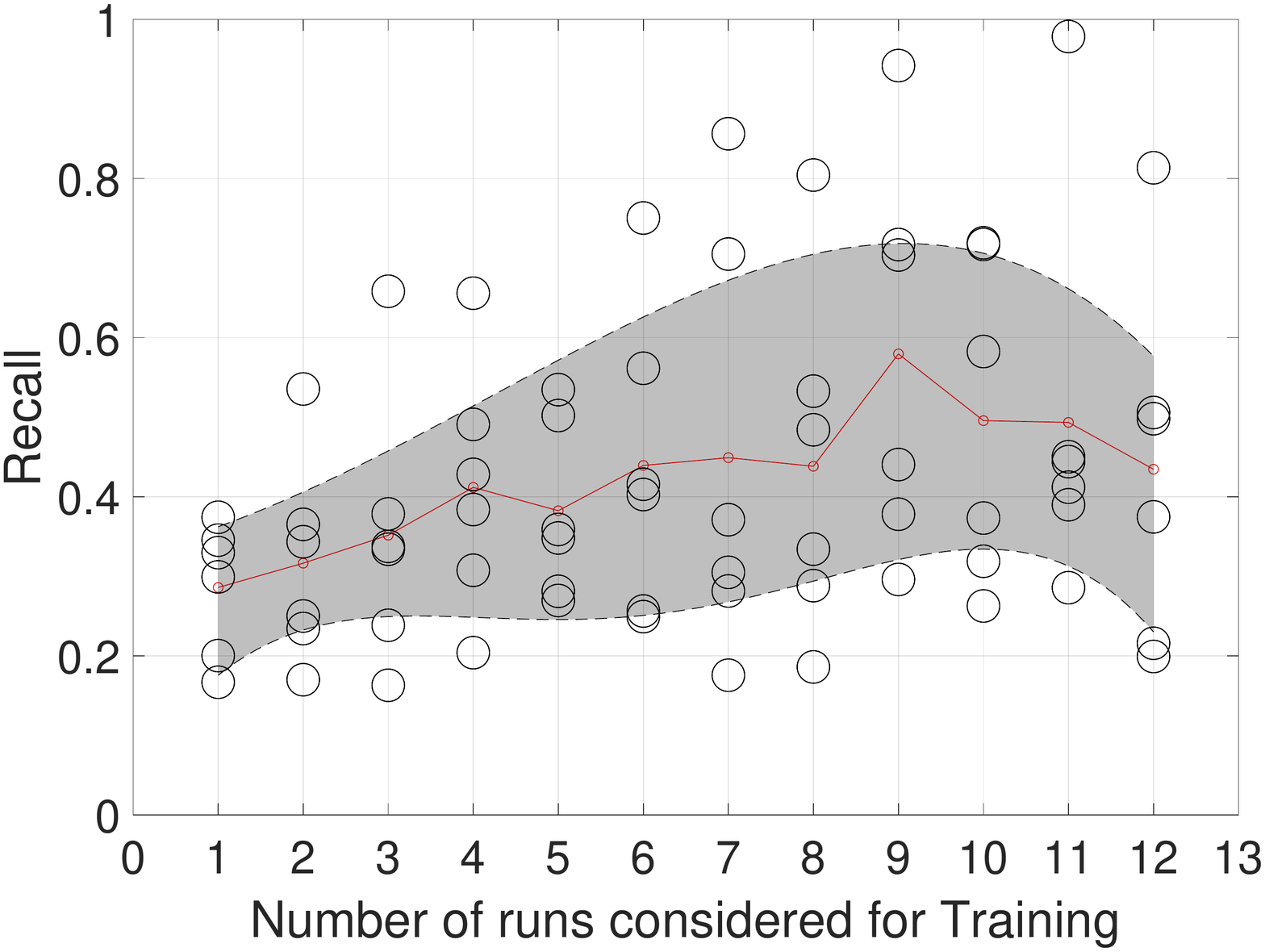}
    \caption*{(b)}  
\end{minipage}
\begin{minipage}{.33\textwidth}
  \centering
  \includegraphics[width=\linewidth]{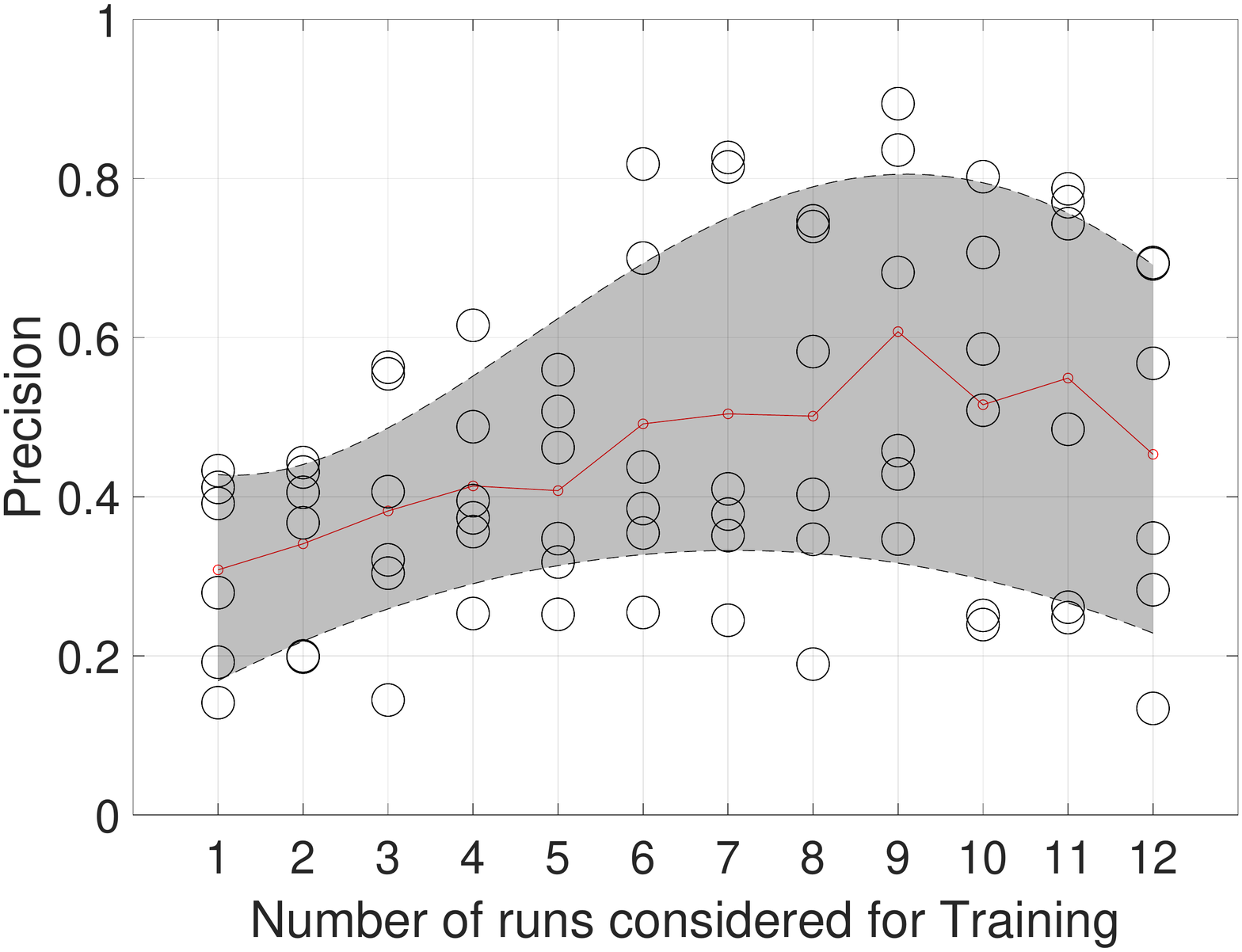}
    \caption*{(c)}
\end{minipage}
\caption{Experiment E3, wired scenario, and raw I-Q samples: the power cycle affects the performance of the classifier in terms of Accuracy (a), Recall (b), and Precision (c). The number of runs (with a power cycle in between) considered for the training process does not affect the performance.}
\label{fig:cable_raw}
\end{figure*}

{\bf power cycle.} We consider the dataset {\em DS1} (from Section~\ref{sec:measurement_collection}), and we run the experiment E1. Given the $13$ runs of {\em DS1}, we randomly choose an increasing subset of the runs for the training process (from $1$ to $12$) and only one run (for all the considered cases) for testing---the testing run is always mutually exclusive with respect to the runs constituting the training set. We repeated this procedure 20 times. Figure~\ref{fig:cable_raw} shows the results of our experiments considering a \ac{CNN} structure similar to the one of~\cite{alshawabka2020} and~\cite{hamdaoui2022}, i.e., {\em ResNet50}, providing as input raw I-Q samples in the form $N \times 1$.
Figure~\ref{fig:cable_raw}(a) shows the accuracy of the classifier when the training process is exposed to an increasing number of runs, i.e., from 1 to 12. For each considered number of runs in the training process, the central mark in the box plots indicates the median value, while the bottom and top edges of the box indicate the percentiles $0.25$ and $0.75$, respectively. The whiskers extend to the most extreme data points not considered outliers, and the outliers are plotted individually using the ``+" marker symbol. The solid red line represents the mean value of the accuracy samples. The accuracy spans approx.  between $0.3$ and $0.6$ (on average) while being characterized by a relatively high variance ($\approx0.3$). These results are consistent with the ones reported in~\cite{alshawabka2020}, where the authors considered the same scenario (wired) and a similar neural network structure. We stress that all the measurements have been taken by power-cycling the radios (both the transmitter and the receiver) every time (E1). Thus, the low performance of the classifier cannot be attributed to the radio channel unpredictability---indeed, we are using a cable, attenuating the dependency on the RF channel propagation. For the sake of completeness, we report recall and precision considering the average value (solid red line), and the region comprised between quantile 10 and 90. Finally, black circles represent the actual outcomes of each specific experiment.
Finally, we highlight that Fig.~\ref{fig:cable_raw} is also taking into account experiment E3. Indeed, since the 13 runs of {\em DS1} are collected over 3 days, i.e., 3 on the first day, 5 on the second one, and finally, 5 on the third one, there are high chances to have adjacent measurements (one after the other, like in E3) when considering a high number of runs, e.g., 12 runs. Thus, Fig.~\ref{fig:cable_raw} captures two distinct phenomena: (i) when the number of runs is small (left side of the x-axis), the training and testing are likely performed on datasets that are temporally far away from each other (E1); and (ii) considering the right part of the x-axis, the training and testing are more and more likely to be temporally close each other (E3)---still being separated by a power cycle. Both experiments (E1 and E3) confirm that the power cycle strongly affects the classifier performance.

\begin{figure}
    \centering
    \includegraphics[width=0.8\columnwidth]{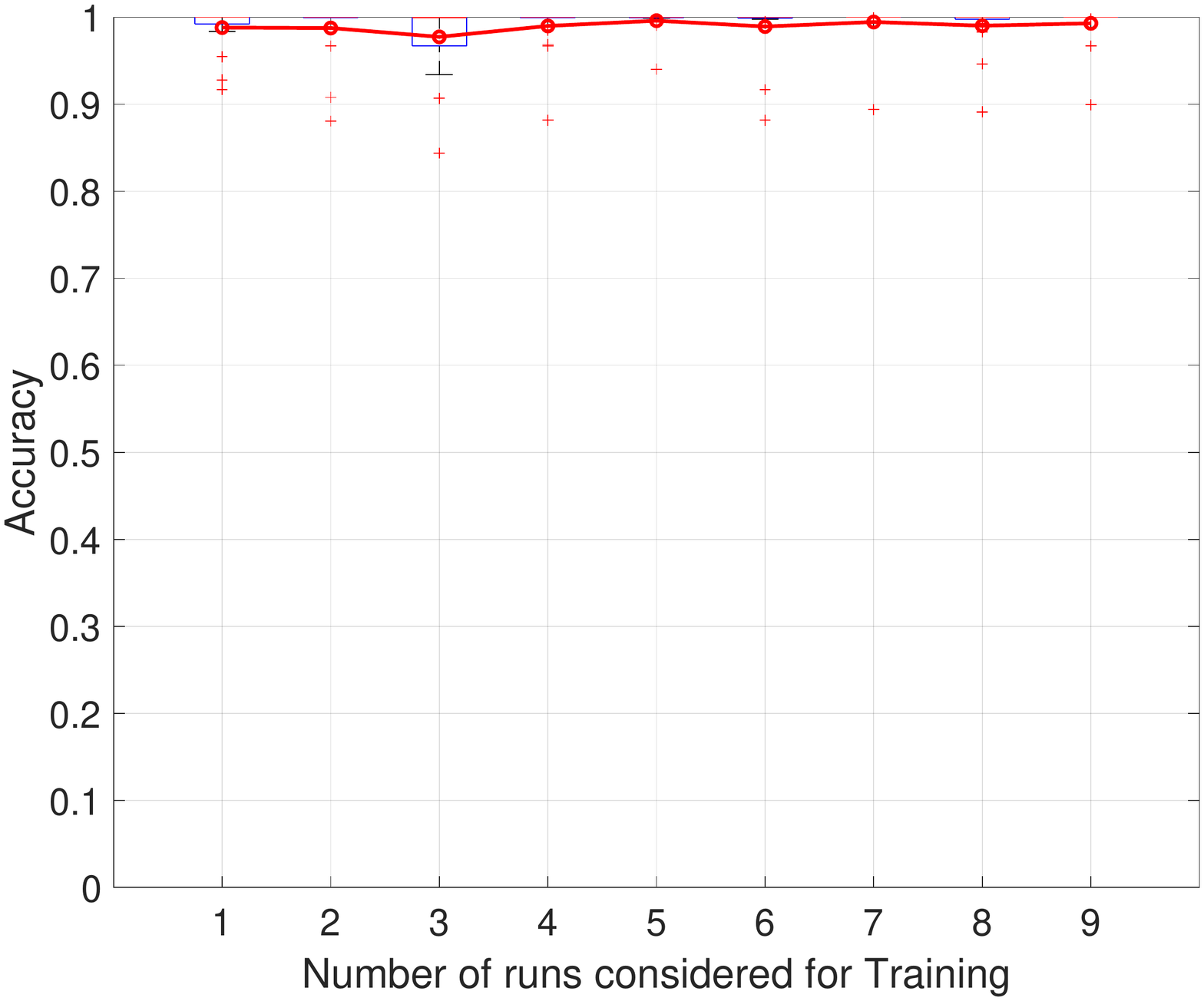}
    \caption{Experiment E4, wired scenario, and raw I-Q samples. The absence of power cycles boosts the performance of the classifier although the runs have been taken on different days. Note how the number of runs considered in the training process does not affect the performance of the classifier.}
    \label{fig:accuracy_3days}
\end{figure}

{\bf No power cycle.} We now consider {\em DS2}, i.e., the long measurements over 3 days, and the same methodology as before. We run experiments E1 and E2 over the long measurements to investigate if the performance of the classifier is affected by either the power cycle (absent) or the temporal distance between two chunks extracted from the measurements. To this aim, we split the 3-day measurements into $10$ chunks, obtaining a total of $60$ chunks. Starting from 6 measures characterized by the same receiver and 6 different transmitters (recall the general set-up from Fig.~\ref{fig:hw_setup}), we randomly selected one or more chunks (up to 9) for training and one for testing (not belonging to the training set). Figure~\ref{fig:accuracy_3days} shows the accuracy (quantiles $0.25$, $0.75$, median, and outliers) of the classifier based on the {\em resnet-50} network. 
\begin{figure}
    \centering
    \includegraphics[width=0.7\columnwidth]{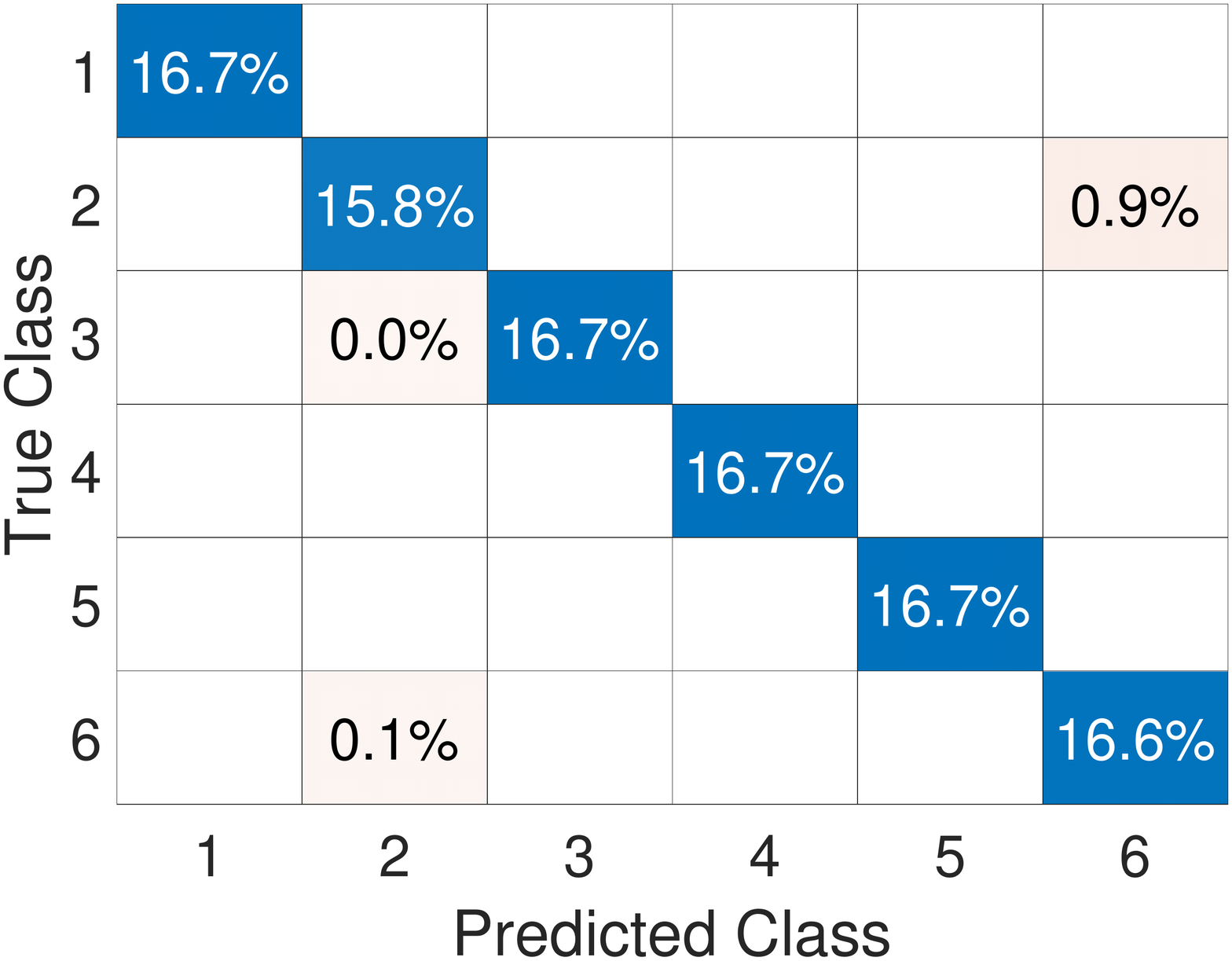}
    \caption{Experiment E4, wired scenario, and raw I-Q samples. The confusion matrix shows outstanding performance for all the considered devices.}
    \label{fig:accuracy_3days_cm}
\end{figure}
The differences between Fig.~\ref{fig:cable_raw}(a) and Fig.~\ref{fig:accuracy_3days} are striking: given the same scenario, i.e., a wired link between the transmitter and the receiver, the different performance of the classifier are due to how the measurements have been collected. If the measurements adopted for training and testing are separated by a power cycle, the performances of the classifier are negatively affected (Fig.~\ref{fig:cable_raw}(a)); conversely, if the training and testing datasets are not separated by a power cycle, the fingerprint is consistent and can be identified with overwhelming probability (Fig.~\ref{fig:accuracy_3days}). Finally, Fig.~\ref{fig:accuracy_3days_cm} provides the normalized confusion matrix considering all the experiments performed for Fig.~\ref{fig:accuracy_3days}: it is worth noting that only about 1\% of the samples are misclassified.

{\bf Wrap-up.} Although we acknowledge that the radio channel variations affect the fingerprinting process, we showed that the power-cycling of the radio plays a major role, as well. Indeed, while common knowledge assumes that low classification accuracy is due to different channel conditions between the measurement adopted for training and the one for testing (Experiment E1 in Fig.~\ref{fig:DAT_effect}), we proved that two consecutive measurements are affected by the same effect when a power cycle is performed in-between (Experiment E3 in Fig.~\ref{fig:power_cycle})---we observed this phenomenon by considering a wired link, so being independent of the multipath fading. Moreover, we also confute the common assumption that measurements taken during the same day (Experiment E2 in Fig.~\ref{fig:DAT_effect}) do not suffer the \acs{DAT} effect, due to the high correlation of the experienced radio channels. Indeed, by considering experiment E4 in Fig.~\ref{fig:power_cycle}, the two chunks (M7 and M8) are separated by a long time but still allow a high classification accuracy due to the absence of the power cycle between M7 and M8. 

\subsection{\acs{DAT} mitigation}
\label{sec:DAT_mitigation}
\begin{figure*}
    \centering
    \includegraphics[trim=0cm 2cm 0cm 2cm,width=\textwidth]{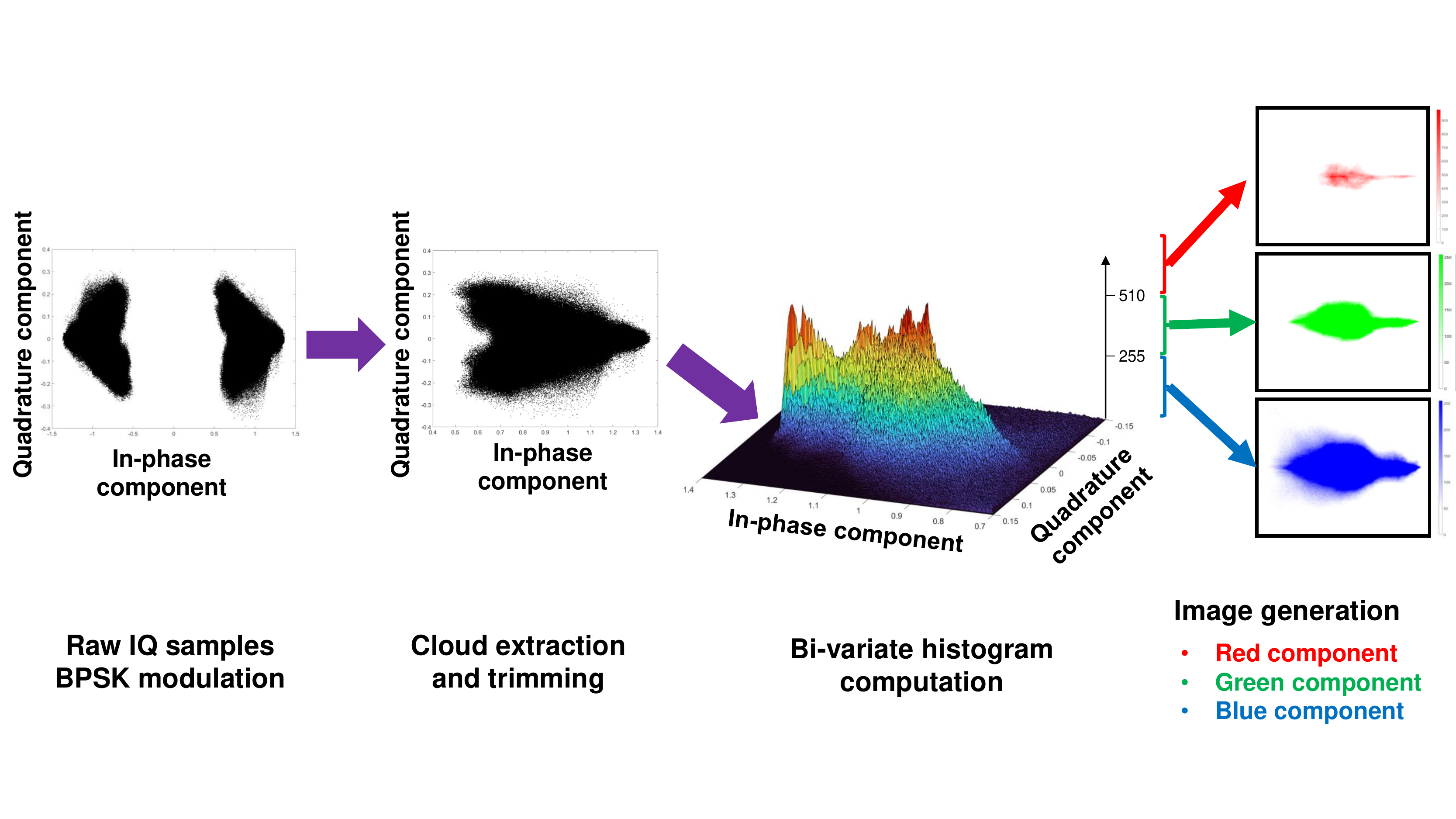}
    \caption{Our solution to mitigate the \acs{DAT} effect: I-Q samples are pre-processed to generate images.}
    \label{fig:algo_images}
\end{figure*}

In this section, we introduce a possible solution to mitigate the \acs{DAT} effect through a dedicated pre-processing of the I-Q samples. Our intuition is rooted in the observation that raw I-Q samples are too noisy, even when considering controlled scenarios (wired). Therefore, we consider a de-noising technique based on the spatial and temporal averaging of the I-Q samples. 
Inspired by~\cite{oligeri2022}, we propose a modified pre-processing technique as depicted in Fig.~\ref{fig:algo_images}. Our methodology considers raw I-Q samples as input from the modulated signal, in this case, BPSK. The raw I-Q samples are mainly organized in two clouds, i.e., one cloud represents the bits equal to 1, while the other cloud represents the zeros---similar considerations can be extended for more complex modulation schemes. The first step consists of defining a chunk size to process the I-Q samples: we consider a chunk size equal to $100,000$ I-Q samples, already taken into account in~\cite{oligeri2022}. The subsequent step differentiates from the reference approach by splitting the original chunk into two chunks containing the left and right clouds, respectively. We trim each cloud and compute a bi-variate histogram, by dividing the I-Q plane into $224 \times 224$ tiles (the size of the images to be used as input to the {\em resnet50} network). Then for each tile, we count the number of received I-Q samples. Contrary to~\cite{oligeri2022}, we consider three layers for the generation of the images, i.e., one layer for each primary colour component (red, green, and blue). Therefore, assuming an image constituted by a three-layer matrix, i.e., $[224 \times 224 \times 3]$ (one layer for each primary colour), and the pixel value between 0 and 255, we assign each value of the tile through the following rule.
\begin{itemize}
    \item $0 \le x_T \le 255$, then $p_R = 0, p_G = 0, p_B = x_T$,
    \item $256 \le x_T \le 511$, then $p_R = 0, p_G = x_T - 255, p_B = 255$,
    \item $x_T > 511$, then , then $p_R = x_T - 510, p_G = 255, p_B = 255$,
\end{itemize}
where $x_T$ represents the value of the tile from the bi-variate histogram, while $p_R, p_G$ and $p_B$ are the pixel values, i.e., red, green and blue, respectively. Finally, we observe that if $x_T > 767$, it is clipped to 767---this issue can also be controlled by properly adjusting the chunk size. We stress that the boundaries of $x_{T}$ are derived from pixel values. Since the pixel value is comprised between 0 and 255, we attribute the colors red ([0, 255]), green (256 + [0, 255]), and blue (256*2 + [0, 255]) as functions of such boundaries. In fact, the output of the bivariate histogram is assigned to a color according to the interval defined previously.
Figure~\ref{fig:algo_images} summarizes the image generation process considering the three image's components, i.e., red, green and blue. Note that the mentioned modification compared to the solution in~\cite{oligeri2022} improves the ability of the \ac{CNN} to adapt to multiple power cycles of the devices. Moreover, as previously demonstrated~\cite{oligeri2022}, converting raw I-Q samples to images is a smart way to average input data over time and space. Previous works have already proved that the (unavoidable) loss of information caused by such techniques does not significantly affect the classifier performance while it mitigates the noise. Indeed, random noise can be filtered out by averaging over time and space: I-Q samples are considered per groups (tiles) in the bivariate histogram. 

\begin{figure*}
\centering
\begin{minipage}{.33\textwidth}
  \centering
  \includegraphics[width=\linewidth]{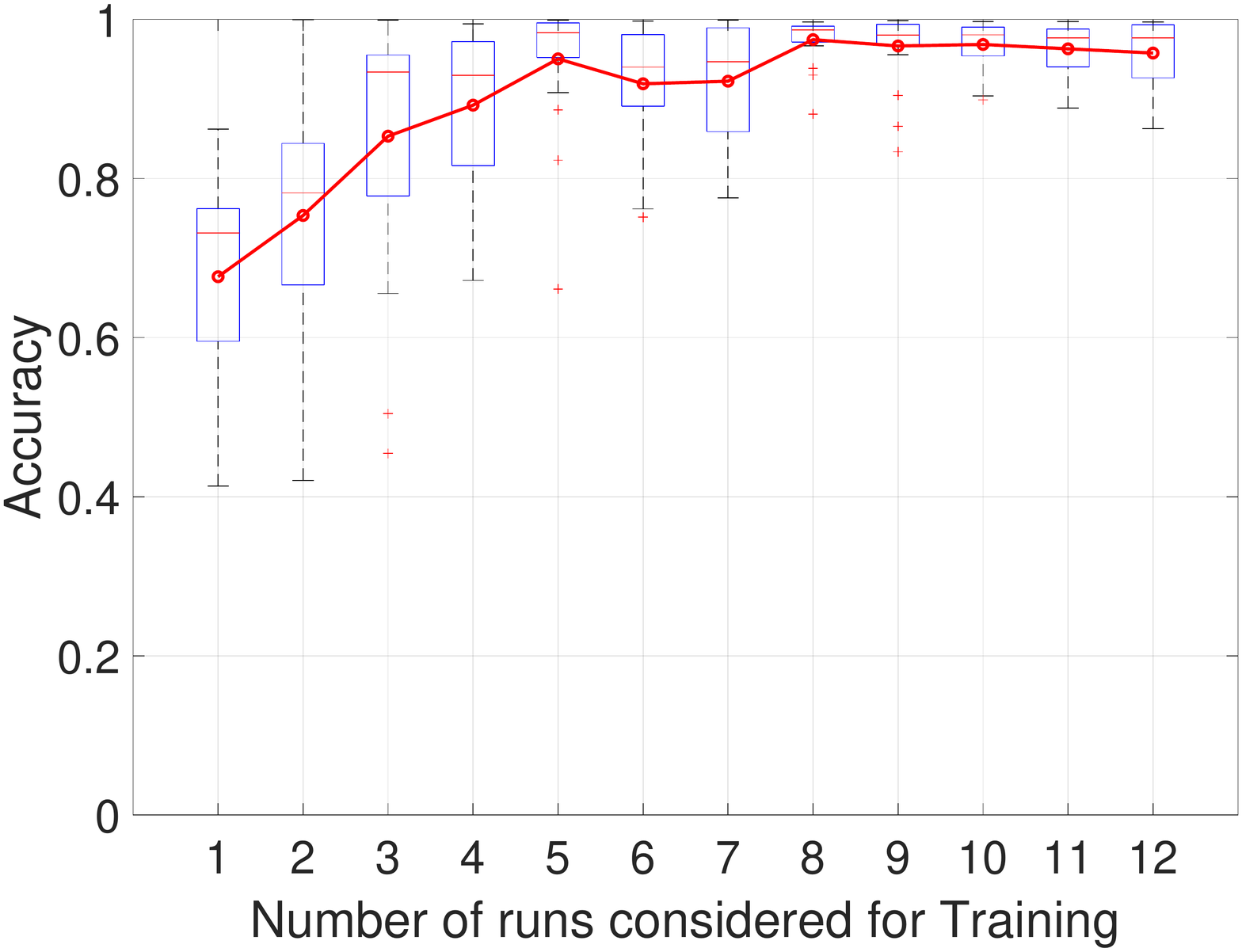}
  \caption*{(a)}
\end{minipage}%
\begin{minipage}{.33\textwidth}
  \centering
  \includegraphics[width=\linewidth]{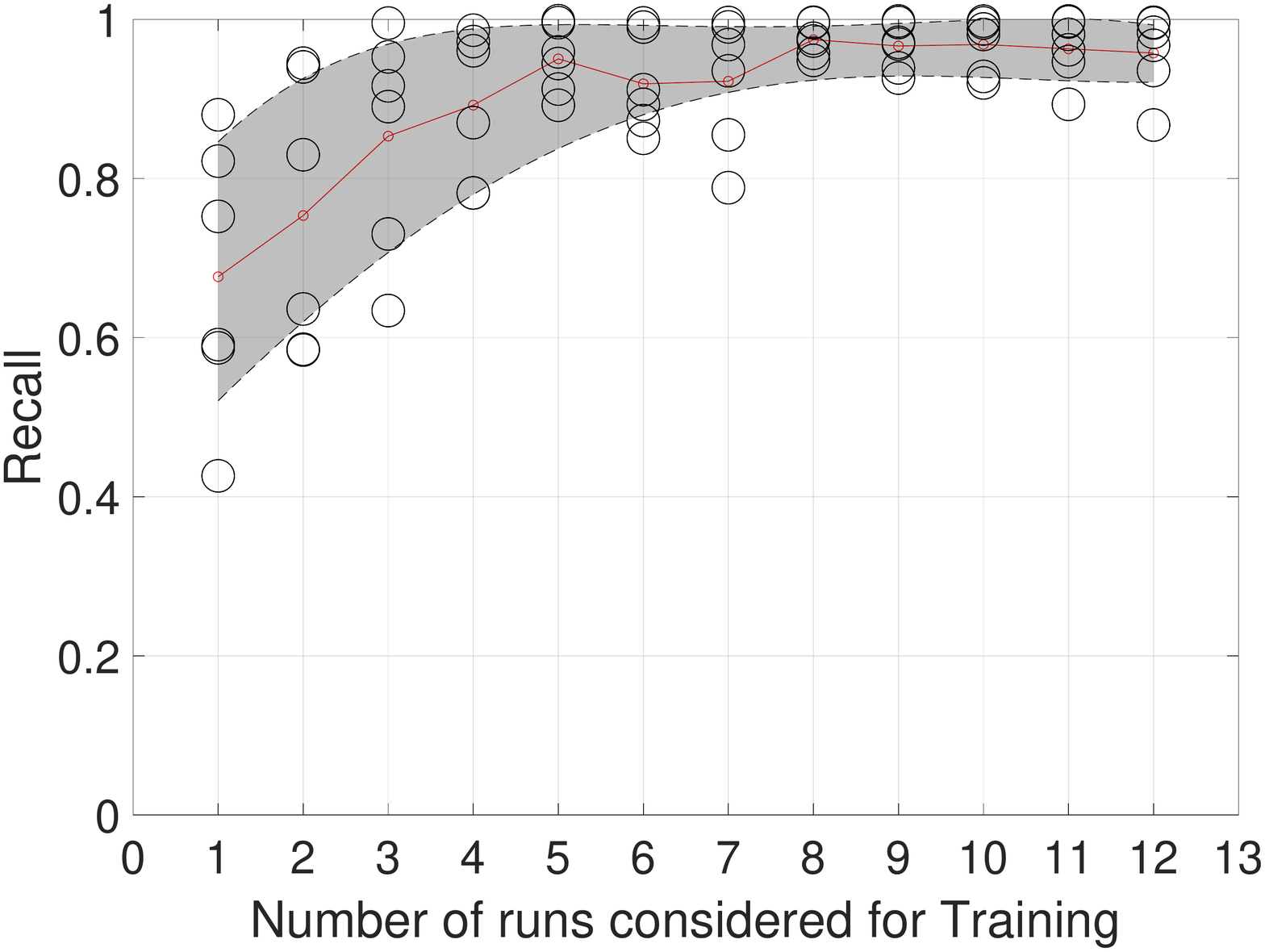}
    \caption*{(b)}
\end{minipage}
\begin{minipage}{.33\textwidth}
  \centering
  \includegraphics[width=\linewidth]{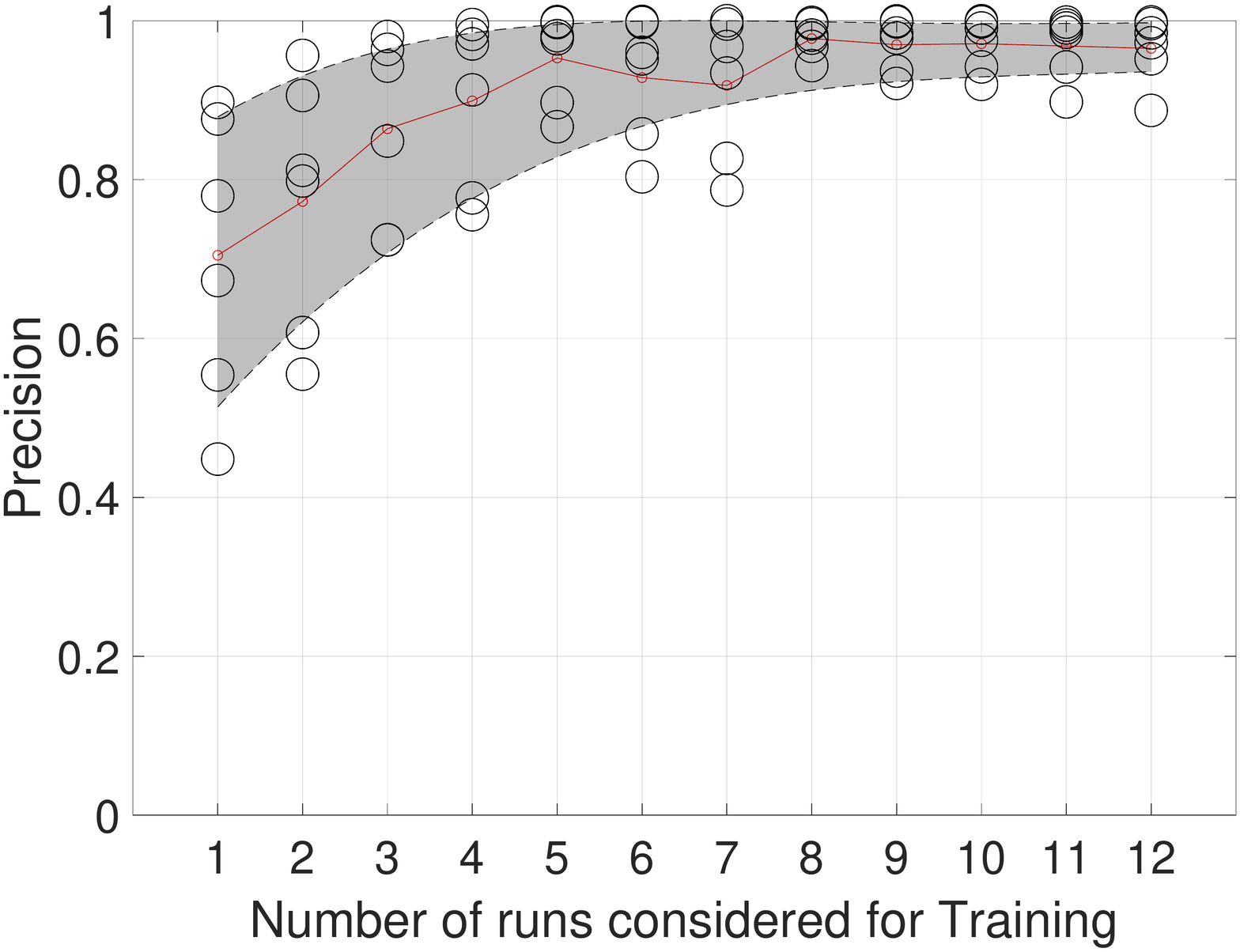}
    \caption*{(c)}
\end{minipage}

\caption{Experiment E3, wired scenario, and raw images: the power cycle affects the performance of the classifier in terms of Accuracy (a), Recall (b), and Precision (c) when the number of runs (with a power cycle in-between) is low as per Fig.~\ref{fig:cable_raw}. When the number of runs increases, the performance is much better ($>0.9$ on average) when using images as input of the classifier.}

\label{fig:cable_images}
\end{figure*}

{\bf Power Cycle.} We now reconsider the Dataset 1 ({\em DS1}) from Sect.~\ref{sec:measurement_collection} and experiments E1 and E3 from Fig.~\ref{fig:DAT_effect} and Fig.~\ref{fig:power_cycle}, respectively. Moreover, we repeated the same experiments as we did before (Fig.~\ref{fig:cable_raw}) by only considering the different procedure presented in Fig.~\ref{fig:algo_images}, i.e., we considered the images as input to the CNN in place of raw I-Q samples. 
Figure~\ref{fig:cable_images} shows the accuracy, recall and precision associated with our tests. We considered an increasing number of randomly chosen runs, from 1 to 12, for the training process, and we tested each configuration $20$ times. For the testing, instead, we considered only one run (random and disjoint from the training set). The performance of the proposed classification methodology is much better compared to the ones of raw I-Q samples adopted by~\cite{alshawabka2020} (recall Fig.~\ref{fig:cable_raw}). Indeed, when training with $5$ randomly chosen runs of measurements, the accuracy is higher than 0.9 while it is about 0.35 when considering raw I-Q samples.  Moreover, exposing the model to more and more runs significantly increases the performance---this is not happening when considering raw I-Q samples. Finally, recall and precision metrics, reported in Figs.~\ref{fig:cable_images}(b) and (c), confirm the quality of our proposed classification algorithm when considering false negatives and false positives, respectively.
\subsection{Real scenario: radio link}
\label{sec:radio_link}

\begin{figure}
    \centering
    \includegraphics[width=\columnwidth]{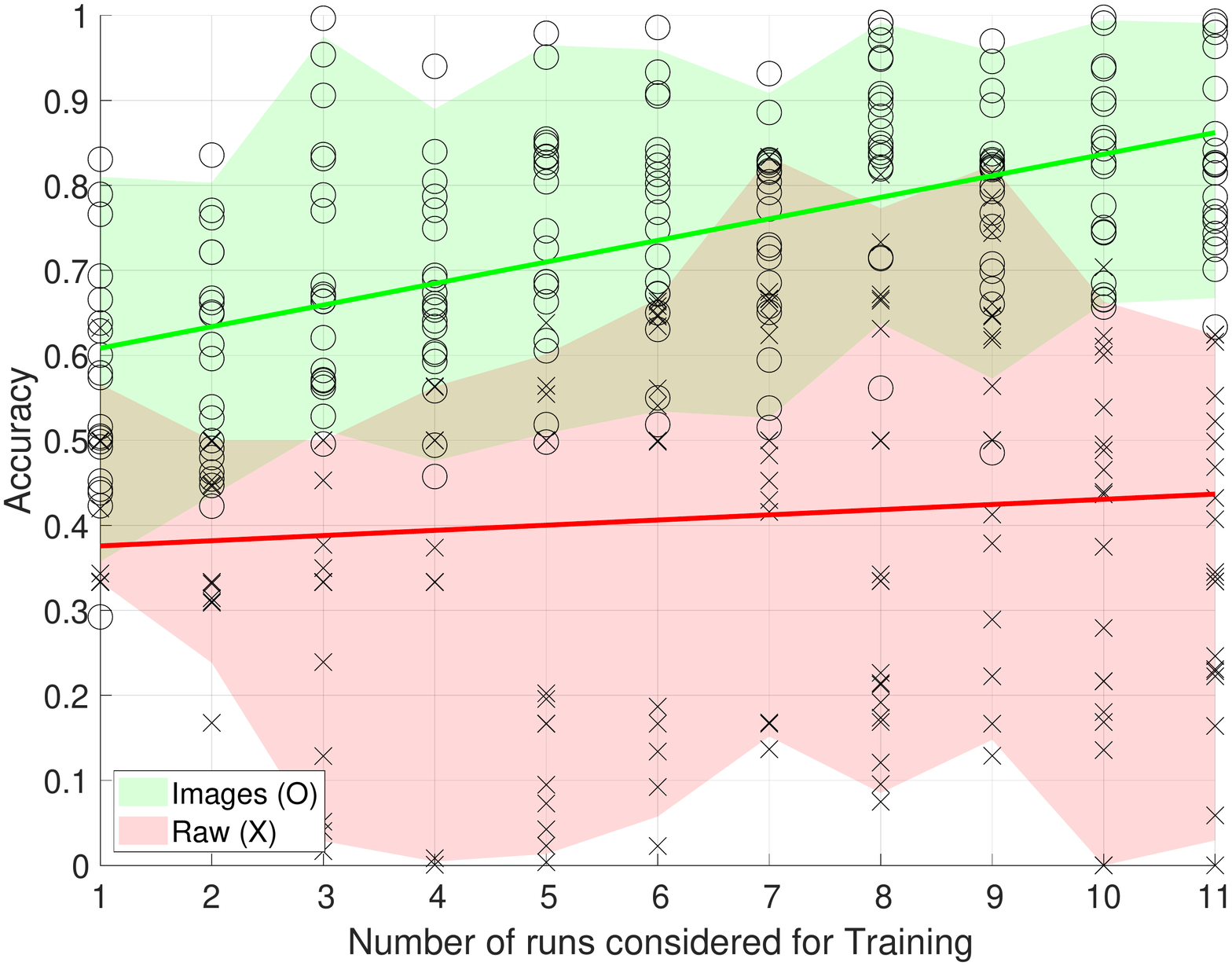}
    \caption{Comparison of raw I-Q samples and images on measurements taken on a wireless link with power cycle.}
    \label{fig:wireless_Im_Raw}
\end{figure}

In this section, we consider a real scenario constituted by the measurement set-up considered in Fig.~\ref{fig:hw_setup}, with a wireless link working at the frequency $f_0= 900MHz$. In this context, we compare the performance of the classifiers considering raw I-Q samples (\cite{alshawabka2020}) and images from dataset {\em DS3}, as described in Sect.~\ref{sec:measurement_collection}. We recall that the measurements in {\em DS3} have been taken to expose as much as possible the \acs{DAT} effect: indeed, each measurement lasts for $5$ minutes and it is collected at random for $4$ days, with several power cycles in-between.
Figure~\ref{fig:wireless_Im_Raw} shows the comparison between the raw I-Q samples and the images when considering {\em resnet50} and dataset {\em DS3}. We adopted the same methodology described before, selecting from 1 to 11 runs for training and a random run for testing, disjoint from the training set. Each black circle and cross in Fig.~\ref{fig:wireless_Im_Raw} represents the outcome of a training/testing process for the images and raw I-Q samples, respectively. The shaded green and red areas refer to the quantiles $0.2$ and $0.8$ calculated on the measured accuracy for both the images and raw I-Q samples. Finally, the solid red and green lines interpolate the accuracy outcomes from the raw I-Q samples and the images, respectively. 
We observe that the performance of our proposed image-based technique outperforms those of raw I-Q samples, i.e., the green shaded area is always well on top of the red one, with a minor overlapping area in the region comprised between 1 and 10 runs. We also highlight the variance associated with the accuracy: raw I-Q samples are characterized by a very high variance ($\approx 0.6$) independently of the number of considered runs. This is a critical issue for the repeatability of the experiments when adopting raw I-Q samples: in many cases, single runs based on raw I-Q samples might experience high accuracy, e.g., higher than $0.8$, leading to exceptional claims about the feasibility of using such a technique for fingerprinting. However, a systematic replication of the training/testing procedure exposes the issue of flat (and poor) performance, which is (almost) independent of the training set size.
Images (shaded green area) behave much better. The average accuracy spans between $\approx0.6$ and $\approx0.85$, depending on the training set size, and our results indicate that extending the training process to a higher number of runs might increase the performance even more. Moreover, we observe that the variance associated with the results is still an issue: even assuming the best configuration (high number of runs), the variance associated with the accuracy is $\approx0.2$, although it decreases when more data is considered for training. In our vision, such a variance might be still relatively high for many real-life applications, requiring smaller confidence. At the same time, we observe that the usage of our technique based on images significantly improves the performance of the RFF process, taking a significant step toward the deployment of RFF techniques in the wild.
Finally, we acknowledge that the \acs{DAT} phenomenon still affects the classification process even when pre-processing I-Q samples into images. In this context, our work shows that techniques based on raw I-Q samples analysis are less robust to power-cycling, whereas solutions based on images generated from I-Q samples have the potential to overcome such an issue, showing promising results. For the sake of completeness, we also report the performance of other DL networks with higher complexity, which require a longer training time. We considered 10 runs and two networks, i.e., {\em inceptionv3} and {\em inceptionresnetv2}, and experienced an average accuracy of 0.89 and 0.89 with variances of 0.013 and 0.010, respectively. Thus, further research is likely required in this area to increase accuracy while reducing associated variance. However, to the best of our knowledge, this manuscript is the first to perform a systematic analysis of the \ac{DAT} effect and demonstrate experimentally the impact of devices' power cycle on the accuracy of \ac{RFF} techniques.
\section{Wrap-up and Discussion}
\label{sec:discussion}

Following the in-depth investigation described in Sect.~\ref{sec:the_day_after_tomorrow_effect}, the \ac{DAT} effect can be re-defined and summarized as the twist of performance affecting the RFF process when the samples used for the training and testing datasets do not belong to the same measurement. A common (and accepted) explanation involves the wireless channel variability, i.e., different measurement times involve a different radio channel, thus causing different performance. The investigation carried out in this work proves this to be a partial (and in some cases ineffective) explanation of the problem.
Although we acknowledge the impact of channel variability, we identified the power cycle of the radios as the main cause of performance loss. This phenomenon is independent of the status of the radio channel and detrimental to the identification of the transmitter, also considering measurements very close in time, and thus experiencing the same channel conditions. Although some authors have already observed it (e.g., \cite{alshawabka2020} and~\cite{hamdaoui2022}), to the best of our knowledge, no one has previously provided a detailed analysis of the \ac{DAT} effect, shedding a light on its causes and possible mitigation strategies.
In our manuscript, we first reproduced the \acs{DAT} effect in a controlled scenario (a wired link between the transmitter and the receiver), by power-cycling the radios under test. Indeed, the power cycle changes the fingerprint enough to significantly reduce the accuracy of the identification process---halving its accuracy in many cases. The \acs{DAT} effect disappears when considering the same measurement set-up, but taking chunks of data from a long measurement not being affected by a power cycle. In this scenario, standard techniques (adopting raw I-Q samples) easily achieve an accuracy close to $1$. These findings show that the \ac{DAT} effect exists independently of the wireless channel, providing new information on the current state of the art.
The causes behind the fingerprint's change are likely linked to the software re-initialization affecting the SDRs after a power cycle. Indeed, at a high-level, SDRs involve mainly two blocks: (i) the FPGA and (ii) the radio-frequency (RF) module. The former (FPGA module) implements the network interface drivers, the data flow command and control, the decimation/interpolation, and finally, the digital up/down conversion (DUC/DDC), while the latter (RF module) implements the analogue transmission of the signal. An important intermediate block between the FPGA and the RF module performs the analogue-to-digital (ADC)/digital-to-analogue (DAC) conversion of the signal. We believe that the power cycle might affect the internal parameters of those blocks, changing the behavior of the radio at the physical-layer.


{\bf DAT analysis.} The DAT effect refers to the change of the \ac{RFF} of a device over time, leading to a significant performance drop of the classifier when training and testing on measurements taken across different days. The DAT effect sums up different factors, such as different signal processing techniques~\cite{brik2008}, non-linear characteristics of power amplifiers that depend on the average output~\cite{kwon2010}, heat dissipation and device temperature~\cite{polak2015}, and clock skews~\cite{zanetti2010}. Nevertheless, we proved that the devices' power cycle significantly affects the performance as well---this being a major cause, since in our analysis we excluded channel impairments and local oscillator drifts (the radios have been calibrated before each measurement), while making our analysis and the associated parameters consistent with other recent work. Finally, we believe that the software nature of the \ac{SDR}---being used as the receiver in all related works---can introduce random artifacts into the data collection process every time they are power cycled, thus preventing transmitter identification.

{\bf Limitations.} Our analysis involved only specific radios, that is, USRP Ettus X310, and we cannot infer the impact of the \acs{DAT} effect on other devices. However, our analysis is in line with the findings of the authors in~\cite{alshawabka2020}, where they observed the same phenomenon considering 20 USRP N210 and USRP X310. Similarly, the authors in~\cite{hamdaoui2022} observed the phenomenon considering USRP B210 and 25 Pycom Lora-enabled devices. Therefore, we are confident in the general validity of our results. Finally, we are confident that the \ac{DAT} effect has been observed by many other researchers, but since its cause can be easily attributed to the wrong source (channel variability), it did not receive the attention it deserves.

{\bf Robustness to Spoofing Attacks.} Recent scientific contributions, such as~\cite{oligeri2022}, experimentally investigated the robustness of \ac{RF} fingerprinting techniques against spoofing attacks. They found that, independently of the chosen approach (being either raw I-Q samples or images), an attacker can defeat \ac{RF} fingerprinting solutions if it injects less than $\frac{N}{2}$ samples into the flow of the samples, being $N$ the number of I-Q samples adopted to create an image. \\
The findings reported in our manuscript are orthogonal to these considerations in spoofing attacks. Indeed, in line with the current literature and considering the approach presented in Sect.~\ref{sec:DAT_mitigation}, adversaries injecting less than $\frac{N}{2}$ samples ($N=100,000$) are likely not to be detected. However, it should be noted that assuming the same number of I-Q samples of competing solutions, such as~\cite{alshawabka2020} and~\cite{hamdaoui2022}, our proposed \ac{DAT} mitigation strategy achieves significantly higher accuracy. In fact, our proposed solution requires fewer I-Q samples to generate reliable \ac{RF} fingerprinting solutions, thus reducing the maximum number of I-Q samples that an adversary can inject without being detected. In this context, we also highlight that, at the time of this writing, adversaries have been able to inject only fully-formed packets (either replayed or fully-crafted), while on-the-fly modification of I-Q samples emitted by legitimate transmitters is out of the technological capabilities of currently-available hardware. Therefore, \ac{RF} fingerprinting solutions are considered effective and robust mainly when applied to high-bandwidth communication links, such as WiFi. For other communication technologies, for example, \ac{IoT} protocols involving low data-rates, \ac{RF} fingerprinting should only be considered as an additional layer of security and coupled with other cryptography-based solutions.

\section{Conclusion and Future Work}
\label{sec:conclusion}
A critical aspect in the successful deployment of \acl{RFF} for physical-layer device authentication is to improve its reliability and robustness. In this paper, we have identified and characterized a new factor that affects the performance of RFF, that is, the power cycle of the devices under test.  Our results show that power cycling the radios has a negative impact on the performance of RFF. To mitigate such an issue, we used a technique based on pre-processing raw I-Q samples collected at the PHY layer. The samples are transformed into images and fed into a \emph{ResNet} \acl{CNN}. Through a comprehensive performance evaluation, we have shown that such an approach mitigates the impact of both power-cycling and wireless channel fluctuations, resulting in an average classification accuracy of $0.85$. This is a significant improvement compared to the average accuracy of approximately 0.5 obtained using state-of-the-art techniques based on raw I-Q samples. However, the variance observed in our results (smaller than competing solutions) suggests that a larger dataset may be necessary to achieve more reliable testing results. Our future work will focus on evaluating the performance of our technique with additional communication technologies, devices, networks, and its associated parameters, as well as its robustness to distance and noise, while also considering the power cycle of either the transmitter or the receiver (independently).
\begin{acks}
Authors would like to thank the anonymous Reviewers for their constructive comments on the article. This publication was made possible by the GSRA7-1-0510-20045 and NPRP12C-0814-190012-SP165 awards from Qatar National Research Fund (a member of Qatar Foundation). This work has also been partially supported by the INTERSECT project, Grant No. NWA.1162.18.301, funded by the Netherlands Organisation for Scientific Research (NWO). The content herein are solely the responsibility of the authors.
\end{acks}
\bibliographystyle{ACM-Reference-Format}
\balance
\bibliography{main}


\begin{thebibliography}{39}


\ifx \showCODEN    \undefined \def \showCODEN     #1{\unskip}     \fi
\ifx \showDOI      \undefined \def \showDOI       #1{#1}\fi
\ifx \showISBNx    \undefined \def \showISBNx     #1{\unskip}     \fi
\ifx \showISBNxiii \undefined \def \showISBNxiii  #1{\unskip}     \fi
\ifx \showISSN     \undefined \def \showISSN      #1{\unskip}     \fi
\ifx \showLCCN     \undefined \def \showLCCN      #1{\unskip}     \fi
\ifx \shownote     \undefined \def \shownote      #1{#1}          \fi
\ifx \showarticletitle \undefined \def \showarticletitle #1{#1}   \fi
\ifx \showURL      \undefined \def \showURL       {\relax}        \fi
\providecommand\bibfield[2]{#2}
\providecommand\bibinfo[2]{#2}
\providecommand\natexlab[1]{#1}
\providecommand\showeprint[2][]{arXiv:#2}

\bibitem[Abbas et~al\mbox{.}(2021)]%
        {abbas2021_nca}
\bibfield{author}{\bibinfo{person}{Sohail Abbas}, \bibinfo{person}{Qassim
  Nasir}, \bibinfo{person}{Douae Nouichi}, \bibinfo{person}{Mohamed
  Abdelsalam}, \bibinfo{person}{Manar Abu~Talib}, \bibinfo{person}{Omnia
  Abu~Waraga}, {et~al\mbox{.}}} \bibinfo{year}{2021}\natexlab{}.
\newblock \showarticletitle{{Improving Security of the Internet of Things via
  RF fingerprinting based device identification system}}.
\newblock \bibinfo{journal}{\emph{Neural Computing and Applications}}
  \bibinfo{volume}{33}, \bibinfo{number}{21} (\bibinfo{year}{2021}),
  \bibinfo{pages}{14753--14769}.
\newblock


\bibitem[Al-Shawabka et~al\mbox{.}(2021)]%
        {al2021deeplora}
\bibfield{author}{\bibinfo{person}{Amani Al-Shawabka}, \bibinfo{person}{Philip
  Pietraski}, \bibinfo{person}{Sudhir~B Pattar}, \bibinfo{person}{Francesco
  Restuccia}, {and} \bibinfo{person}{Tommaso Melodia}.}
  \bibinfo{year}{2021}\natexlab{}.
\newblock \showarticletitle{DeepLoRa: Fingerprinting LoRa devices at scale
  through deep learning and data augmentation}. In
  \bibinfo{booktitle}{\emph{Proceedings of the Twenty-second International
  Symposium on Theory, Algorithmic Foundations, and Protocol Design for Mobile
  Networks and Mobile Computing}}. \bibinfo{pages}{251--260}.
\newblock


\bibitem[Al-Shawabka et~al\mbox{.}(2020)]%
        {alshawabka2020}
\bibfield{author}{\bibinfo{person}{Amani Al-Shawabka},
  \bibinfo{person}{Francesco Restuccia}, \bibinfo{person}{Salvatore D’Oro},
  \bibinfo{person}{Tong Jian}, \bibinfo{person}{Bruno Costa~Rendon},
  \bibinfo{person}{Nasim Soltani}, \bibinfo{person}{Jennifer Dy},
  \bibinfo{person}{Stratis Ioannidis}, \bibinfo{person}{Kaushik Chowdhury},
  {and} \bibinfo{person}{Tommaso Melodia}.} \bibinfo{year}{2020}\natexlab{}.
\newblock \showarticletitle{Exposing the Fingerprint: Dissecting the Impact of
  the Wireless Channel on Radio Fingerprinting}. In
  \bibinfo{booktitle}{\emph{IEEE INFOCOM 2020 - IEEE Conference on Computer
  Communications}} (Toronto, ON, Canada). \bibinfo{publisher}{IEEE Press},
  \bibinfo{pages}{646–655}.
\newblock
\urldef\tempurl%
\url{https://doi.org/10.1109/INFOCOM41043.2020.9155259}
\showDOI{\tempurl}


\bibitem[Ali et~al\mbox{.}(2019)]%
        {ali2019_access}
\bibfield{author}{\bibinfo{person}{Aysha~M Ali}, \bibinfo{person}{Emre
  Uzundurukan}, {and} \bibinfo{person}{Ali Kara}.}
  \bibinfo{year}{2019}\natexlab{}.
\newblock \showarticletitle{{Assessment of features and classifiers for
  Bluetooth RF fingerprinting}}.
\newblock \bibinfo{journal}{\emph{IEEE Access}}  \bibinfo{volume}{7}
  (\bibinfo{year}{2019}), \bibinfo{pages}{50524--50535}.
\newblock


\bibitem[Bihl et~al\mbox{.}(2016)]%
        {bihl2016_tifs}
\bibfield{author}{\bibinfo{person}{Trevor~J. Bihl}, \bibinfo{person}{Kenneth~W.
  Bauer}, {and} \bibinfo{person}{Michael~A. Temple}.}
  \bibinfo{year}{2016}\natexlab{}.
\newblock \showarticletitle{{Feature Selection for RF Fingerprinting With
  Multiple Discriminant Analysis and Using ZigBee Device Emissions}}.
\newblock \bibinfo{journal}{\emph{IEEE Transactions on Information Forensics
  and Security}} \bibinfo{volume}{11}, \bibinfo{number}{8}
  (\bibinfo{year}{2016}), \bibinfo{pages}{1862--1874}.
\newblock
\urldef\tempurl%
\url{https://doi.org/10.1109/TIFS.2016.2561902}
\showDOI{\tempurl}


\bibitem[Brik et~al\mbox{.}(2008)]%
        {brik2008}
\bibfield{author}{\bibinfo{person}{Vladimir Brik}, \bibinfo{person}{Suman
  Banerjee}, \bibinfo{person}{Marco Gruteser}, {and} \bibinfo{person}{Sangho
  Oh}.} \bibinfo{year}{2008}\natexlab{}.
\newblock \showarticletitle{Wireless Device Identification with Radiometric
  Signatures}. In \bibinfo{booktitle}{\emph{Proceedings of the 14th ACM
  International Conference on Mobile Computing and Networking}} (San Francisco,
  California, USA) \emph{(\bibinfo{series}{MobiCom '08})}.
  \bibinfo{publisher}{Association for Computing Machinery},
  \bibinfo{address}{New York, NY, USA}, \bibinfo{pages}{116–127}.
\newblock
\showISBNx{9781605580968}
\urldef\tempurl%
\url{https://doi.org/10.1145/1409944.1409959}
\showDOI{\tempurl}


\bibitem[Ding et~al\mbox{.}(2018)]%
        {ding2018specific}
\bibfield{author}{\bibinfo{person}{Lida Ding}, \bibinfo{person}{Shilian Wang},
  \bibinfo{person}{Fanggang Wang}, {and} \bibinfo{person}{Wei Zhang}.}
  \bibinfo{year}{2018}\natexlab{}.
\newblock \showarticletitle{Specific emitter identification via convolutional
  neural networks}.
\newblock \bibinfo{journal}{\emph{IEEE Communications Letters}}
  \bibinfo{volume}{22}, \bibinfo{number}{12} (\bibinfo{year}{2018}),
  \bibinfo{pages}{2591--2594}.
\newblock


\bibitem[{DSP StackExchange}(2017)]%
        {sampling_same}
\bibfield{author}{\bibinfo{person}{{DSP StackExchange}}.}
  \bibinfo{year}{2017}\natexlab{}.
\newblock \bibinfo{title}{{Bandwidth with complex sampling}}.
\newblock
  \bibinfo{howpublished}{\url{https://dsp.stackexchange.com/questions/36927/bandwidth-with-complex-sampling}}.
\newblock
\newblock
\shownote{(Accessed: 2023-Sep-28)}.


\bibitem[Elmaghbub and Hamdaoui(2021)]%
        {elmaghbub2021lora}
\bibfield{author}{\bibinfo{person}{Abdurrahman Elmaghbub} {and}
  \bibinfo{person}{Bechir Hamdaoui}.} \bibinfo{year}{2021}\natexlab{}.
\newblock \showarticletitle{{LoRa device fingerprinting in the wild: Disclosing
  RF data-driven fingerprint sensitivity to deployment variability}}.
\newblock \bibinfo{journal}{\emph{IEEE Access}}  \bibinfo{volume}{9}
  (\bibinfo{year}{2021}), \bibinfo{pages}{142893--142909}.
\newblock


\bibitem[{Ettus Research}(2020a)]%
        {ubx}
\bibfield{author}{\bibinfo{person}{{Ettus Research}}.}
  \bibinfo{year}{2020}\natexlab{a}.
\newblock \bibinfo{title}{{UBX160 Daughterboard}}.
\newblock
  \bibinfo{howpublished}{\url{https://www.ettus.com/product/details/UBX160}}.
\newblock
\newblock
\shownote{(Accessed: 2023-Sep-28)}.


\bibitem[{Ettus Research}(2020b)]%
        {ettus}
\bibfield{author}{\bibinfo{person}{{Ettus Research}}.}
  \bibinfo{year}{2020}\natexlab{b}.
\newblock \bibinfo{title}{{USRP X310}}.
\newblock
  \bibinfo{howpublished}{\url{https://www.ettus.com/all-products/x310-kit/}}.
\newblock
\newblock
\shownote{(Accessed: 2023-Sep-28)}.


\bibitem[Gong et~al\mbox{.}(2020)]%
        {gong2020_tifs}
\bibfield{author}{\bibinfo{person}{Jialiang Gong}, \bibinfo{person}{Xiaodong
  Xu}, {and} \bibinfo{person}{Yingke Lei}.} \bibinfo{year}{2020}\natexlab{}.
\newblock \showarticletitle{{Unsupervised specific emitter identification
  method using radio-frequency fingerprint embedded InfoGAN}}.
\newblock \bibinfo{journal}{\emph{IEEE Transactions on Information Forensics
  and Security}}  \bibinfo{volume}{15} (\bibinfo{year}{2020}),
  \bibinfo{pages}{2898--2913}.
\newblock


\bibitem[Gul et~al\mbox{.}(2022)]%
        {9966888}
\bibfield{author}{\bibinfo{person}{Omer~Melih Gul}, \bibinfo{person}{Michel
  Kulhandjian}, \bibinfo{person}{Burak Kantarci}, \bibinfo{person}{Azzedine
  Touazi}, \bibinfo{person}{Cliff Ellement}, {and} \bibinfo{person}{Claude
  D'Amours}.} \bibinfo{year}{2022}\natexlab{}.
\newblock \showarticletitle{Fine-grained Augmentation for RF Fingerprinting
  under Impaired Channels}. In \bibinfo{booktitle}{\emph{2022 IEEE 27th
  International Workshop on Computer Aided Modeling and Design of Communication
  Links and Networks (CAMAD)}}. \bibinfo{pages}{115--120}.
\newblock
\urldef\tempurl%
\url{https://doi.org/10.1109/CAMAD55695.2022.9966888}
\showDOI{\tempurl}


\bibitem[Hamdaoui and Elmaghbub(2022)]%
        {hamdaoui2022}
\bibfield{author}{\bibinfo{person}{Bechir Hamdaoui} {and}
  \bibinfo{person}{Abdurrahman Elmaghbub}.} \bibinfo{year}{2022}\natexlab{}.
\newblock \showarticletitle{Deep-Learning-Based Device Fingerprinting for
  Increased LoRa-IoT Security: Sensitivity to Network Deployment Changes}.
\newblock \bibinfo{journal}{\emph{IEEE Network}} \bibinfo{volume}{36},
  \bibinfo{number}{3} (\bibinfo{year}{2022}), \bibinfo{pages}{204--210}.
\newblock
\urldef\tempurl%
\url{https://doi.org/10.1109/MNET.001.2100553}
\showDOI{\tempurl}


\bibitem[Hanna et~al\mbox{.}(2022)]%
        {hanna2022wisig}
\bibfield{author}{\bibinfo{person}{Samer Hanna}, \bibinfo{person}{Samurdhi
  Karunaratne}, {and} \bibinfo{person}{Danijela Cabric}.}
  \bibinfo{year}{2022}\natexlab{}.
\newblock \showarticletitle{WiSig: A Large-Scale WiFi Signal Dataset for
  Receiver and Channel Agnostic RF Fingerprinting}.
\newblock \bibinfo{journal}{\emph{IEEE Access}}  \bibinfo{volume}{10}
  (\bibinfo{year}{2022}), \bibinfo{pages}{22808--22818}.
\newblock


\bibitem[Jagannath et~al\mbox{.}(2022)]%
        {jagannath2022_arxiv}
\bibfield{author}{\bibinfo{person}{Anu Jagannath}, \bibinfo{person}{Jithin
  Jagannath}, {and} \bibinfo{person}{Prem Sagar Pattanshetty~Vasanth Kumar}.}
  \bibinfo{year}{2022}\natexlab{}.
\newblock \showarticletitle{{A Comprehensive Survey on Radio Frequency (RF)
  Fingerprinting: Traditional Approaches, Deep Learning, and Open Challenges}}.
\newblock \bibinfo{journal}{\emph{arXiv preprint arXiv:2201.00680}}
  (\bibinfo{year}{2022}).
\newblock


\bibitem[Jian et~al\mbox{.}(2021)]%
        {jian2021_tmc}
\bibfield{author}{\bibinfo{person}{Tong Jian}, \bibinfo{person}{Yifan Gong},
  \bibinfo{person}{Zheng Zhan}, \bibinfo{person}{Runbin Shi},
  \bibinfo{person}{Nasim Soltani}, \bibinfo{person}{Zifeng Wang},
  \bibinfo{person}{Jennifer Dy}, \bibinfo{person}{Kaushik Chowdhury},
  \bibinfo{person}{Yanzhi Wang}, {and} \bibinfo{person}{Stratis Ioannidis}.}
  \bibinfo{year}{2021}\natexlab{}.
\newblock \showarticletitle{{Radio Frequency Fingerprinting on the Edge}}.
\newblock \bibinfo{journal}{\emph{IEEE Transactions on Mobile Computing}}
  \bibinfo{volume}{21}, \bibinfo{number}{11} (\bibinfo{year}{2021}),
  \bibinfo{pages}{4078--4093}.
\newblock


\bibitem[Jian et~al\mbox{.}(2020)]%
        {jian2020deep}
\bibfield{author}{\bibinfo{person}{Tong Jian}, \bibinfo{person}{Bruno~Costa
  Rendon}, \bibinfo{person}{Emmanuel Ojuba}, \bibinfo{person}{Nasim Soltani},
  \bibinfo{person}{Zifeng Wang}, \bibinfo{person}{Kunal Sankhe},
  \bibinfo{person}{Andrey Gritsenko}, \bibinfo{person}{Jennifer Dy},
  \bibinfo{person}{Kaushik Chowdhury}, {and} \bibinfo{person}{Stratis
  Ioannidis}.} \bibinfo{year}{2020}\natexlab{}.
\newblock \showarticletitle{Deep learning for RF fingerprinting: A massive
  experimental study}.
\newblock \bibinfo{journal}{\emph{IEEE Internet of Things Magazine}}
  \bibinfo{volume}{3}, \bibinfo{number}{1} (\bibinfo{year}{2020}),
  \bibinfo{pages}{50--57}.
\newblock


\bibitem[Kwon et~al\mbox{.}(2010)]%
        {kwon2010}
\bibfield{author}{\bibinfo{person}{Dae~Hyun Kwon}, \bibinfo{person}{Hao Li},
  \bibinfo{person}{Yuchun Chang}, \bibinfo{person}{Richard Tseng}, {and}
  \bibinfo{person}{Yun Chiu}.} \bibinfo{year}{2010}\natexlab{}.
\newblock \showarticletitle{Digitally Equalized CMOS Transmitter Front-End With
  Integrated Power Amplifier}.
\newblock \bibinfo{journal}{\emph{IEEE Journal of Solid-State Circuits}}
  \bibinfo{volume}{45}, \bibinfo{number}{8} (\bibinfo{year}{2010}),
  \bibinfo{pages}{1602--1614}.
\newblock
\urldef\tempurl%
\url{https://doi.org/10.1109/JSSC.2010.2048140}
\showDOI{\tempurl}


\bibitem[Lathi et~al\mbox{.}(1995)]%
        {lathi}
\bibfield{author}{\bibinfo{person}{B.~P. Lathi}, \bibinfo{person}{Adel~S.
  Sedra}, {and} \bibinfo{person}{M.E.~Van Valkenburg}.}
  \bibinfo{year}{1995}\natexlab{}.
\newblock \bibinfo{booktitle}{\emph{Modern Digital and Analog Communication
  Systems} (\bibinfo{edition}{2nd} ed.)}.
\newblock \bibinfo{publisher}{Oxford University Press, Inc.},
  \bibinfo{address}{USA}.
\newblock
\showISBNx{0195105001}


\bibitem[Merchant et~al\mbox{.}(2018)]%
        {merchant2018deep}
\bibfield{author}{\bibinfo{person}{Kevin Merchant}, \bibinfo{person}{Shauna
  Revay}, \bibinfo{person}{George Stantchev}, {and} \bibinfo{person}{Bryan
  Nousain}.} \bibinfo{year}{2018}\natexlab{}.
\newblock \showarticletitle{Deep learning for RF device fingerprinting in
  cognitive communication networks}.
\newblock \bibinfo{journal}{\emph{IEEE Journal of Selected Topics in Signal
  Processing}} \bibinfo{volume}{12}, \bibinfo{number}{1}
  (\bibinfo{year}{2018}), \bibinfo{pages}{160--167}.
\newblock


\bibitem[Mohanti et~al\mbox{.}(2020)]%
        {mohanti2020airid}
\bibfield{author}{\bibinfo{person}{Subhramoy Mohanti}, \bibinfo{person}{Nasim
  Soltani}, \bibinfo{person}{Kunal Sankhe}, \bibinfo{person}{Dheryta
  Jaisinghani}, \bibinfo{person}{Marco Di~Felice}, {and}
  \bibinfo{person}{Kaushik Chowdhury}.} \bibinfo{year}{2020}\natexlab{}.
\newblock \showarticletitle{AirID: Injecting a custom RF fingerprint for
  enhanced UAV identification using deep learning}. In
  \bibinfo{booktitle}{\emph{GLOBECOM 2020-2020 IEEE Global Communications
  Conference}}. IEEE, \bibinfo{pages}{1--6}.
\newblock


\bibitem[Moreno-Torres et~al\mbox{.}(2012)]%
        {moreno2012unifying}
\bibfield{author}{\bibinfo{person}{Jose~G Moreno-Torres}, \bibinfo{person}{Troy
  Raeder}, \bibinfo{person}{Roc{\'\i}o Alaiz-Rodr{\'\i}guez},
  \bibinfo{person}{Nitesh~V Chawla}, {and} \bibinfo{person}{Francisco
  Herrera}.} \bibinfo{year}{2012}\natexlab{}.
\newblock \showarticletitle{{A Unifying View on Dataset Shift in
  Classification}}.
\newblock \bibinfo{journal}{\emph{Pattern recognition}} \bibinfo{volume}{45},
  \bibinfo{number}{1} (\bibinfo{year}{2012}), \bibinfo{pages}{521--530}.
\newblock


\bibitem[Oligeri et~al\mbox{.}(2022)]%
        {oligeri2022}
\bibfield{author}{\bibinfo{person}{Gabriele Oligeri}, \bibinfo{person}{Savio
  Sciancalepore}, \bibinfo{person}{Simone Raponi}, {and}
  \bibinfo{person}{Roberto Di~Pietro}.} \bibinfo{year}{2022}\natexlab{}.
\newblock \showarticletitle{{PAST-AI: Physical-layer Authentication of
  Satellite Transmitters via Deep Learning}}.
\newblock \bibinfo{journal}{\emph{IEEE Transactions on Information Forensics
  and Security}} (\bibinfo{year}{2022}), \bibinfo{pages}{1--1}.
\newblock
\urldef\tempurl%
\url{https://doi.org/10.1109/TIFS.2022.3219287}
\showDOI{\tempurl}


\bibitem[Polak and Goeckel(2015)]%
        {polak2015}
\bibfield{author}{\bibinfo{person}{Adam~C. Polak} {and}
  \bibinfo{person}{Dennis~L. Goeckel}.} \bibinfo{year}{2015}\natexlab{}.
\newblock \showarticletitle{Identification of Wireless Devices of Users Who
  Actively Fake Their RF Fingerprints With Artificial Data Distortion}.
\newblock \bibinfo{journal}{\emph{IEEE Transactions on Wireless
  Communications}} \bibinfo{volume}{14}, \bibinfo{number}{11}
  (\bibinfo{year}{2015}), \bibinfo{pages}{5889--5899}.
\newblock
\urldef\tempurl%
\url{https://doi.org/10.1109/TWC.2015.2443794}
\showDOI{\tempurl}


\bibitem[Rajendran et~al\mbox{.}(2020)]%
        {rajendran2020injecting}
\bibfield{author}{\bibinfo{person}{Sekhar Rajendran}, \bibinfo{person}{Zhi
  Sun}, \bibinfo{person}{Feng Lin}, {and} \bibinfo{person}{Kui Ren}.}
  \bibinfo{year}{2020}\natexlab{}.
\newblock \showarticletitle{Injecting reliable radio frequency fingerprints
  using metasurface for the Internet of Things}.
\newblock \bibinfo{journal}{\emph{IEEE Transactions on Information Forensics
  and Security}}  \bibinfo{volume}{16} (\bibinfo{year}{2020}),
  \bibinfo{pages}{1896--1911}.
\newblock


\bibitem[Restuccia et~al\mbox{.}(2019)]%
        {restuccia2019deepradioid}
\bibfield{author}{\bibinfo{person}{Francesco Restuccia},
  \bibinfo{person}{Salvatore D'Oro}, \bibinfo{person}{Amani Al-Shawabka},
  \bibinfo{person}{Mauro Belgiovine}, \bibinfo{person}{Luca Angioloni},
  \bibinfo{person}{Stratis Ioannidis}, \bibinfo{person}{Kaushik Chowdhury},
  {and} \bibinfo{person}{Tommaso Melodia}.} \bibinfo{year}{2019}\natexlab{}.
\newblock \showarticletitle{DeepRadioID: Real-time channel-resilient
  optimization of deep learning-based radio fingerprinting algorithms}. In
  \bibinfo{booktitle}{\emph{Proceedings of the Twentieth ACM International
  Symposium on Mobile Ad Hoc Networking and Computing}}.
  \bibinfo{pages}{51--60}.
\newblock


\bibitem[Restuccia et~al\mbox{.}(2021)]%
        {restuccia2021deepfir}
\bibfield{author}{\bibinfo{person}{Francesco Restuccia},
  \bibinfo{person}{Salvatore D’Oro}, \bibinfo{person}{Amani Al-Shawabka},
  \bibinfo{person}{Bruno~Costa Rendon}, \bibinfo{person}{Stratis Ioannidis},
  {and} \bibinfo{person}{Tommaso Melodia}.} \bibinfo{year}{2021}\natexlab{}.
\newblock \showarticletitle{DeepFIR: Channel-Robust Physical-Layer Deep
  Learning Through Adaptive Waveform Filtering}.
\newblock \bibinfo{journal}{\emph{IEEE Transactions on Wireless
  Communications}} \bibinfo{volume}{20}, \bibinfo{number}{12}
  (\bibinfo{year}{2021}), \bibinfo{pages}{8054--8066}.
\newblock


\bibitem[Riyaz et~al\mbox{.}(2018)]%
        {riyaz2018deep}
\bibfield{author}{\bibinfo{person}{Shamnaz Riyaz}, \bibinfo{person}{Kunal
  Sankhe}, \bibinfo{person}{Stratis Ioannidis}, {and} \bibinfo{person}{Kaushik
  Chowdhury}.} \bibinfo{year}{2018}\natexlab{}.
\newblock \showarticletitle{Deep learning convolutional neural networks for
  radio identification}.
\newblock \bibinfo{journal}{\emph{IEEE Communications Magazine}}
  \bibinfo{volume}{56}, \bibinfo{number}{9} (\bibinfo{year}{2018}),
  \bibinfo{pages}{146--152}.
\newblock


\bibitem[Sankhe et~al\mbox{.}(2019)]%
        {sankhe2019oracle}
\bibfield{author}{\bibinfo{person}{Kunal Sankhe}, \bibinfo{person}{Mauro
  Belgiovine}, \bibinfo{person}{Fan Zhou}, \bibinfo{person}{Shamnaz Riyaz},
  \bibinfo{person}{Stratis Ioannidis}, {and} \bibinfo{person}{Kaushik
  Chowdhury}.} \bibinfo{year}{2019}\natexlab{}.
\newblock \showarticletitle{ORACLE: Optimized radio classification through
  convolutional neural networks}. In \bibinfo{booktitle}{\emph{IEEE INFOCOM
  2019-IEEE Conference on Computer Communications}}. IEEE,
  \bibinfo{pages}{370--378}.
\newblock


\bibitem[Shen et~al\mbox{.}(2022)]%
        {shen2022towards}
\bibfield{author}{\bibinfo{person}{Guanxiong Shen}, \bibinfo{person}{Junqing
  Zhang}, \bibinfo{person}{Alan Marshall}, {and} \bibinfo{person}{Joseph~R
  Cavallaro}.} \bibinfo{year}{2022}\natexlab{}.
\newblock \showarticletitle{{Towards scalable and channel-robust radio
  frequency fingerprint identification for LoRa}}.
\newblock \bibinfo{journal}{\emph{IEEE Transactions on Information Forensics
  and Security}}  \bibinfo{volume}{17} (\bibinfo{year}{2022}),
  \bibinfo{pages}{774--787}.
\newblock


\bibitem[Shen et~al\mbox{.}(2021)]%
        {shen2021radio}
\bibfield{author}{\bibinfo{person}{Guanxiong Shen}, \bibinfo{person}{Junqing
  Zhang}, \bibinfo{person}{Alan Marshall}, \bibinfo{person}{Linning Peng},
  {and} \bibinfo{person}{Xianbin Wang}.} \bibinfo{year}{2021}\natexlab{}.
\newblock \showarticletitle{{Radio Frequency Fingerprint Identification for
  LoRa using Deep Learning}}.
\newblock \bibinfo{journal}{\emph{IEEE Journal on Selected Areas in
  Communications}} \bibinfo{volume}{39}, \bibinfo{number}{8}
  (\bibinfo{year}{2021}), \bibinfo{pages}{2604--2616}.
\newblock


\bibitem[Soltani et~al\mbox{.}(2020)]%
        {soltani2020more}
\bibfield{author}{\bibinfo{person}{Nasim Soltani}, \bibinfo{person}{Kunal
  Sankhe}, \bibinfo{person}{Jennifer Dy}, \bibinfo{person}{Stratis Ioannidis},
  {and} \bibinfo{person}{Kaushik Chowdhury}.} \bibinfo{year}{2020}\natexlab{}.
\newblock \showarticletitle{More is better: Data augmentation for
  channel-resilient RF fingerprinting}.
\newblock \bibinfo{journal}{\emph{IEEE Communications Magazine}}
  \bibinfo{volume}{58}, \bibinfo{number}{10} (\bibinfo{year}{2020}),
  \bibinfo{pages}{66--72}.
\newblock


\bibitem[Soltanieh et~al\mbox{.}(2020)]%
        {soltanieh2020_jrid}
\bibfield{author}{\bibinfo{person}{Naeimeh Soltanieh}, \bibinfo{person}{Yaser
  Norouzi}, \bibinfo{person}{Yang Yang}, {and} \bibinfo{person}{Nemai~Chandra
  Karmakar}.} \bibinfo{year}{2020}\natexlab{}.
\newblock \showarticletitle{{A Review of Radio Frequency Fingerprinting
  Techniques}}.
\newblock \bibinfo{journal}{\emph{IEEE Journal of Radio Frequency
  Identification}} \bibinfo{volume}{4}, \bibinfo{number}{3}
  (\bibinfo{year}{2020}), \bibinfo{pages}{222--233}.
\newblock


\bibitem[Wang and Gan(2022)]%
        {wang2022_tccn}
\bibfield{author}{\bibinfo{person}{Weidong Wang} {and} \bibinfo{person}{Lu
  Gan}.} \bibinfo{year}{2022}\natexlab{}.
\newblock \showarticletitle{{Radio Frequency Fingerprinting Improved by
  Statistical Noise Reduction}}.
\newblock \bibinfo{journal}{\emph{IEEE Transactions on Cognitive Communications
  and Networking}} (\bibinfo{year}{2022}).
\newblock


\bibitem[Yan et~al\mbox{.}(2022)]%
        {yan2022rrf}
\bibfield{author}{\bibinfo{person}{Wenqing Yan}, \bibinfo{person}{Thiemo
  Voigt}, {and} \bibinfo{person}{Christian Rohner}.}
  \bibinfo{year}{2022}\natexlab{}.
\newblock \showarticletitle{RRF: A Robust Radiometric Fingerprint System that
  Embraces Wireless Channel Diversity}. In
  \bibinfo{booktitle}{\emph{Proceedings of the 15th ACM Conference on Security
  and Privacy in Wireless and Mobile Networks}}. \bibinfo{pages}{85--97}.
\newblock


\bibitem[Yu et~al\mbox{.}(2019)]%
        {yu2019robust}
\bibfield{author}{\bibinfo{person}{Jiabao Yu}, \bibinfo{person}{Aiqun Hu},
  \bibinfo{person}{Guyue Li}, {and} \bibinfo{person}{Linning Peng}.}
  \bibinfo{year}{2019}\natexlab{}.
\newblock \showarticletitle{A robust RF fingerprinting approach using
  multisampling convolutional neural network}.
\newblock \bibinfo{journal}{\emph{IEEE Internet of Things Journal}}
  \bibinfo{volume}{6}, \bibinfo{number}{4} (\bibinfo{year}{2019}),
  \bibinfo{pages}{6786--6799}.
\newblock


\bibitem[Zanetti et~al\mbox{.}(2010)]%
        {zanetti2010}
\bibfield{author}{\bibinfo{person}{Davide Zanetti}, \bibinfo{person}{Boris
  Danev}, {and} \bibinfo{person}{Srdjan Capkun}.}
  \bibinfo{year}{2010}\natexlab{}.
\newblock \showarticletitle{Physical-Layer Identification of UHF RFID Tags}. In
  \bibinfo{booktitle}{\emph{Proceedings of the Sixteenth Annual International
  Conference on Mobile Computing and Networking}} (Chicago, Illinois, USA)
  \emph{(\bibinfo{series}{MobiCom '10})}. \bibinfo{publisher}{Association for
  Computing Machinery}, \bibinfo{address}{New York, NY, USA},
  \bibinfo{pages}{353–364}.
\newblock
\showISBNx{9781450301817}
\urldef\tempurl%
\url{https://doi.org/10.1145/1859995.1860035}
\showDOI{\tempurl}


\bibitem[Zhang et~al\mbox{.}(2022)]%
        {zhang2022_commlett}
\bibfield{author}{\bibinfo{person}{Zhen Zhang}, \bibinfo{person}{Aiqun Hu},
  \bibinfo{person}{Wei Xu}, \bibinfo{person}{Jiabao Yu}, {and}
  \bibinfo{person}{Yang Yang}.} \bibinfo{year}{2022}\natexlab{}.
\newblock \showarticletitle{{An Artificial Radio Frequency Fingerprint
  Embedding Scheme for Device Identification}}.
\newblock \bibinfo{journal}{\emph{IEEE Communications Letters}}
  \bibinfo{volume}{26}, \bibinfo{number}{5} (\bibinfo{year}{2022}),
  \bibinfo{pages}{974--978}.
\newblock
\urldef\tempurl%
\url{https://doi.org/10.1109/LCOMM.2022.3148037}
\showDOI{\tempurl}


\end{thebibliography}

\end{document}